\documentclass[12pt]{article}

\RequirePackage[colorlinks,citecolor=blue,urlcolor=blue]{hyperref}
\RequirePackage[OT1]{fontenc}
\RequirePackage{amsthm,amssymb,amsfonts,graphicx,subfigure,caption,bm,fullpage,wrapfig,setspace}

\usepackage{amsmath,amssymb,enumerate}
 \usepackage{float}
\usepackage{algpseudocode,algorithm,algorithmicx}
\usepackage{mathrsfs}
\usepackage{multirow}
\usepackage{forest, adjustbox}
\usepackage{bbm}
\usepackage{lscape}
\usetikzlibrary{calc}
\usepackage{relsize}

\tikzset{fontscale/.style = {font=\relsize{#1}}
    }
\numberwithin{equation}{section}
\theoremstyle{plain}
                          
\theoremstyle{remark}         
\theoremstyle{definition} 

\newcommand\A{\mathcal{A}}

\newcommand\C{\mathcal{C}}

\newcommand\T{\mathcal{T}}
\newcommand\G{\mathcal{G}}
\newcommand\U{\mathcal{U}}
\newcommand\J{\mathcal{J}}

\newcommand\mI{\mathcal{I}}
\newcommand\mL{\mathcal{L}}

\newcommand\mS{\mathcal{S}}

\newcommand\bxi{\bm{\xi}}
\newcommand\bphi{\bm{\phi}}

\renewcommand\S{\mathcal{S}}

\newcommand\bc{\bm{c}}

\newcommand\bx{\bm{x}}
\newcommand\by{\bm{y}}
\newcommand\bz{\bm{z}}

\newcommand\bX{\bm{X}}
\newcommand\bY{\bm{Y}}

\newcommand\brho{\bm\rho}

\newcommand\bbet{\bm\beta}

\newcommand\bpi{\boldsymbol{\pi}}
\newcommand{\RN}[1]{%
  \textup{\uppercase\expandafter{\romannumeral#1}}%
}

\newcommand\bzero{\bm{0}}

\newcommand\bi{\begin{itemize}}
\newcommand\ei{\end{itemize}}

 \allowdisplaybreaks

\newcommand{\beginsupplement}{%
        \setcounter{table}{0}
        \renewcommand{\thetable}{S\arabic{table}}%
        \setcounter{figure}{0}
        \renewcommand{\thefigure}{S\arabic{figure}}%
        \renewcommand{\theequation}{S\arabic{equation}}
}

\def\woMR#1{\w@MR#1MR#1MR\relax}%
\def\w@MR#1MR#2MR#3\relax{#2}

\def\@MR#1 #2\relax#3{%
 \href{http://www.ams.org/mathscinet-getitem?mr=#1}%
 {\MRfixed{#3}}}%

\def\MRfixed{MR\woMR}%

\usepackage{natbib}
\usepackage{varioref}		
\labelformat{section}{Section~#1}
\labelformat{figure}{Figure~#1}
\labelformat{table}{Table~#1}	

\title{Bayesian graphical compositional regression for microbiome data}
\author{Jialiang Mao \and Yuhan Chen \and Li Ma}
\date{\textit{Duke University, Durham, NC} \\ \hspace{1mm} \\ \today}
\begin{document}
\maketitle

\vspace{-3mm}
\begin{abstract}
\doublespacing  An important task in microbiome studies is to test the existence of and give characterization to differences in the microbiome composition across groups of samples. Important challenges of this problem include the large within-group heterogeneities among samples and the existence of potential confounding variables that, when ignored, increase the chance of false discoveries and reduce the power for identifying true differences. {\color{black} We propose a probabilistic framework to overcome these issues by combining three ideas: (i) a phylogenetic tree-based decomposition of the cross-group comparison problem into a series of local tests, (ii) a graphical model that links the local tests to allow information sharing across taxa, and (iii) a Bayesian testing strategy that incorporates covariates and integrates out the within-group variation, avoiding potentially unstable point estimates. We derive an efficient inference algorithm based on numerical integration and junction-tree message passing,} conduct extensive simulation studies to investigate the performance of our approach, {\color{black}and compare it to state-of-the-art methods} in a number of representative settings. We then apply our method to the American Gut data to analyze the association of dietary habits and human's gut microbiome composition in the presence of covariates, and illustrate the importance of incorporating covariates in microbiome cross-group comparison.

\end{abstract}

\doublespacing

\section{Introduction}
\label{sec:intro}
The human microbiome is the community of numerous microbes that inhabit the human body. Understanding the microbiome can provide insights into various aspects of human health. For example, diseases such as obesity and Type 2 diabetes have been shown to be related to the gut microbiome \citep{turnbaugh2006obesity, qin2012metagenome}. 
Next generation sequencing technologies provide ways of profiling the microbiome. This is typically achieved through either shotgun sequencing on the entire genomes of microbes, or through targeting a signature gene---the 16S ribosomal RNA (rRNA) gene that provides barcodes of species identity. The 16S rRNA gene of the bacteria in the samples is sequenced and the sequences are clustered into operational taxonomic units (OTUs) using preprocessing pipelines such as QIIME \citep{caporaso2010qiime}. Traditionally, OTUs are used as representatives of the species at 97\% similarity level {\color{black}and their counts (defined as the counts of sequences in each OTU cluster) form the basis of analyzing the composition of the human microbiome \citep{li2015microbiome}. 
More recently, \cite{callahan2017exact} proposed to use amplicon sequence variants (ASVs) to achieve more precise characterization of species. Our methodology herein applies to both OTUs and ASVs though we will use ``OTUs'' to refer to the unit of species.  }

One important task of microbiome studies is to compare the composition of the microbial community of groups of subjects \citep{hildebrandt2009high, wu2011linking, qin2012metagenome, david2014diet}. 
Various methods have been proposed for comparing two groups of data samples and their underlying distributions, ranging from the classic $t$-test 
to numerous recently developed methods (see for example \cite{holmes2015two}, \cite{soriano2017probabilistic} and \cite{chen2017new}). However, these generic approaches are either inapplicable or severely underpowered when the samples are OTU counts. 

There are several features of the OTU counts that make this two-sample-problem challenging: (i) high dimensionality: the number of OTUs in the study is often large, (ii) {\color{black}``rare biosphere'' \citep{sogin2006microbial} or sparsity: the total OTU count of a sample is dominated by a few OTUs, with others having counts closed or equal to zero}, (iii) complex covariance structure: the correlations among counts of different OTUs are complicated, and (iv) large overdispersions: the counts of samples in a same group often show large within-group heterogeneities. {\color{black}Fortunately, along with the OTU counts, one can construct a phylogenetic tree that encodes the evolutionary relationships of these OTUs and collect abundant covariate information about the host participant of each sample. In this paper, we propose a testing method, called Bayesian graphical compositional regression (BGCR), that utilizes such information to tackle the aforementioned challenges.} 

{\color{black}BGCR aims to effectively account for the specific features of microbiome data by combining three techniques---(i) a decomposition of the multinomial likelihood along the phylogenetic tree, (ii) graphical models , and (iii) the Bayesian testing framework. Broadly speaking, our approach falls into a stream of works based on the Dirichlet-multinomial model (DM) \citep{la:2012} that puts a Dirichlet prior on the multinomial parameters to resolve the large overdispersions in OTU counts. Recent developments along the line use the phylogenetic tree---which summarizes the evolutionary relationship among the OTUs and thus serves as a proxy to their functional relationship---to enrich the model construction. Specifically, \cite{wang2017dirichlet} adopt the Dirichlet tree multinomial model (DTM) in which OTU counts are aggregated along the phylogenetic tree and a DM model is introduced to characterize how the OTU counts on each internal node of the tree are distributed into its child nodes. BGCR 
adopts a similar decomposition to transform the original testing problem into multiple testing involving a collection of node-specific tests along the phylogenetic tree.}

{\color{black}Instead of treating the node-specific tests independently, we use a bottom-up graphical structure to introduce dependency among them. 
Our motivation is that some of these tests may have poor statistical power due to the limited number of samples and relevant OTU counts. The dependency structure allows effective information sharing among the node-specific problems, thereby improving the power of the tests. Our choice of a bottom-up autoregressive specification comes from the observation that} compositional differences are often observed to cluster into ``chains'' along the phylogenetic tree \citep{tang:2017}. Such chains can arise in two ways: either due to a structural constraint of DTM---a cross-group difference at any particular node in the phylogenetic tree induces (weaker) differences in its ancestors, or as a result of the functional relatedness of OTUs in the phylogenetic tree.

{\color{black}The decomposition scheme and the graphical structure are merged with the Bayesian testing framework to give fast and interpretable inference through a suite of computational techniques including numerical integration and message passing. On the one hand, exact posterior summaries are available through a recursive algorithm that does not require Monte Carlo simulations; on the other, the testing results are given a full probabilistic characterization with the uncertainty quantified in a coherent manner.} In addition, we show that the Bayesian test results in substantially improved power, likely due to the way the within-group variability is dealt with---integrated out rather than estimated as in existing approaches.

{\color{black}Besides the phylogenetic tree, covariate information is also used to assist the modeling of the microbome composition in recent works {\color{black} focusing} on prediction \citep{tang2017mixed} and selection of the covariates with significant associations to the OTU counts \citep{xia2013logistic, wang2017dirichlet, wadsworth2017integrative,  ren2017bayesian, grantham2017mimixf}. In contrast, little attention has been paid to the cross-group comparison scenario
, where incorporating covariates is even more crucial since many microbiome studies are observational and the effects of 
 unadjusted confounders could invalidate the testing results by creating false positives. As a starting point, BGCR allows a node-specific regression adjustment for important covariates. This adjustment can be incorporated in the inference mechanism with little extra computational burden and is able to reduce the number of false positives effectively, as we will illustrate in our numerical examples.}

{\color{black}In summary, in comparison to existing approaches, the contribution of BGCR are three-fold: (i) it allows a Bayesian framework for testing cross-group differences in microbiome composition along a phylogenetic structure; (ii) it provides a principled probabilistic modeling framework for borrowing information across taxa, thereby enhancing the power for detecting cross-group differences; and (iii) it uses a principled way to incorporate additional covariates in the Bayesian testing setting while maintaining the computational efficiency of the proposed framework.} 

To close the introduction, we connect BGCR to some relevant references in the DM literature stream. Using DM, \cite{la:2012} proposed a generalized Wald-type test statistic based on method-of-moments estimates of the Dirichlet parameters. Due to the large number of OTUs, this test has large degrees of freedom and is usually underpowered. One simple attempt to alleviate this problem is to aggregate OTUs from the same genus and study their composition at the genus level as suggested by \cite{chen2013variable}. However, one could also aggregate the OTUs to other levels in the hierarchy of biological classification, such as the family or order level, yet different levels of aggregation can result in inconsistent testing results \citep{tang:2017}. Similar to BGCR, \cite{tang:2017} proposed the PhyloScan test, which also decomposes the original problem to a series of node-specific tests along the phylogenetic tree and tries to incorporate dependencies among these tests. BGCR and PhyloScan differ in the way they deal with these underlying dependencies. PhyloScan introduces dependencies through a scanning procedure by adding up the test statistics in triplets of neighboring nodes while BGCR considers a probabilistic graphical model that describes the dependencies in a generative manner. The latter allows more flexible borrowing of information beyond neighboring triplets along with a fully probabilistic interpretation.

In \ref{sec:method}, we describe our method for testing and characterizing the cross-group differences of OTU compositions. In \ref{sec:num_exam}, four representative simulation scenarios are considered to evaluate the performance of the proposed method. An application of the method to the American Gut data is shown in \ref{sec:app}. \ref{sec:discussion} concludes.


\section{Method}
\label{sec:method}


\subsection{Data and background}
 \label{subsec:DM}
 
In this section, {\color{black}we set up some notations and briefly review DM and DTM for OTU compositions. The microbiome dataset we work with contains three parts: OTU counts in each sample, a phylogenetic tree over the OTUs, and a set of covariates for each sample.}

\textit{OTU counts.} Consider a microbiome dataset with OTU counts of two groups of subjects on $K$ OTUs denoted by $\Omega=\{\text{OTU}_1,\text{OTU}_2,\ldots, \text{OTU}_K \} = \{ \omega_1,\omega_2,\ldots,\omega_K\}$. Let $n_i$ be the number of samples in group $i$, $i=0, 1$. For the $j$-th sample in group $i$, $j=1,\ldots, n_i$,  let $\by_{ij}=(y_{ij1},\ldots,y_{ijK})$ be the vector of its OTU counts, where $y_{ijl}$ is the number of the $l$-th OTU in this sample for $l=1,\ldots,K$. Moreover, let $N_{ij}=\sum^{K}_{l=1}y_{ijl}$ be the total number of OTU counts in that sample. For simplicity, let $\bY$ be the OTU counts of all the samples. {\color{black}\ref{table:toy} illustrate the OTU counts for a certain group of samples.}

\textit{Phylogenetic tree.} Let $\T$ be a {\color{black} rooted full binary} phylogenetic tree that describes the evolutionary relations of the $K$ OTUs. Let $\mI$ denote the set of its internal nodes (i.e., non-leaf nodes). We denote each node $A$ of $\T$ by the set of its descendant OTUs. For example, $A=\{\omega_l \}$, $1\leq l\leq K$ represents a leaf of $\T$ that contains a single OTU $\omega_i$; $A=\Omega$ is the root containing all the OTUs. Since the tree is full, each internal node has exactly two children. For $A\in \mI$, let $A_l, A_r$ be the left and right children of $A$. If $A\not=\Omega$, let $A_p$ be its parent {\color{black} and $A_s$ be its sibling (i.e., the node in $\T$ that has the same parent as $A$)}. {\color{black}  If $A$ is a leaf, for the $j$-th sample in group $i$, we let $y_{ij}(A)=y_{ijl}$ be the count of the OTU that $A$ represents; if $A\in\mI$, we recursively define the count in $A$ to be the aggregation of the counts in its two children: $y_{ij}(A)=y_{ij}(A_l) + y_{ij}(A_r)$. Equivalently, $y_{ij}(A)=\sum_{\{l:\omega_l\in A\}}y_{ijl}$. \ref{fig:toy} shows an example of a phylogenetic tree over $6$ OTUs. In this example, $A=\{\omega_3,\omega_4,\omega_5,\omega_6\}$, $y_{ij}(A)=y_{ij3}+y_{ij4}+y_{ij5} + y_{ij6}.$ }

\vspace{8mm}

\begin{minipage}{\textwidth}   
  \begin{minipage}[b]{0.45\textwidth}
    \centering
    \begin{tabular}{  r | r r r r | r  }
  \hline
id   &  $\omega_1$ & $\omega_2$& $\cdots$ & $\omega_K$ & Sum \\\hline
1 & $y_{i11}$ & $y_{i12}$ & $\cdots$ & $y_{i1K}$  & $N_{i1}$\\
2 &  $y_{i21}$ & $y_{i22}$ & $\cdots$ & $y_{i2K}$ & $N_{i2}$\\
$\vdots$  & $\vdots$ & $\vdots$ & $\ddots$ & $\vdots$ & $\vdots$ \\
$n_i$ &  $y_{in_i1}$ & $y_{in_i2}$ & $\cdots$ & $y_{in_iK}$ & $N_{in_i}$ \\
   \hline
  \end{tabular}
      \captionof{table}{{\color{black}OTU counts for the $i$-th group.}}
        \label{table:toy}
  \end{minipage}
     \hfill
   \begin{minipage}[b]{0.49\textwidth}
    \centering
    \includegraphics[width = 6cm]{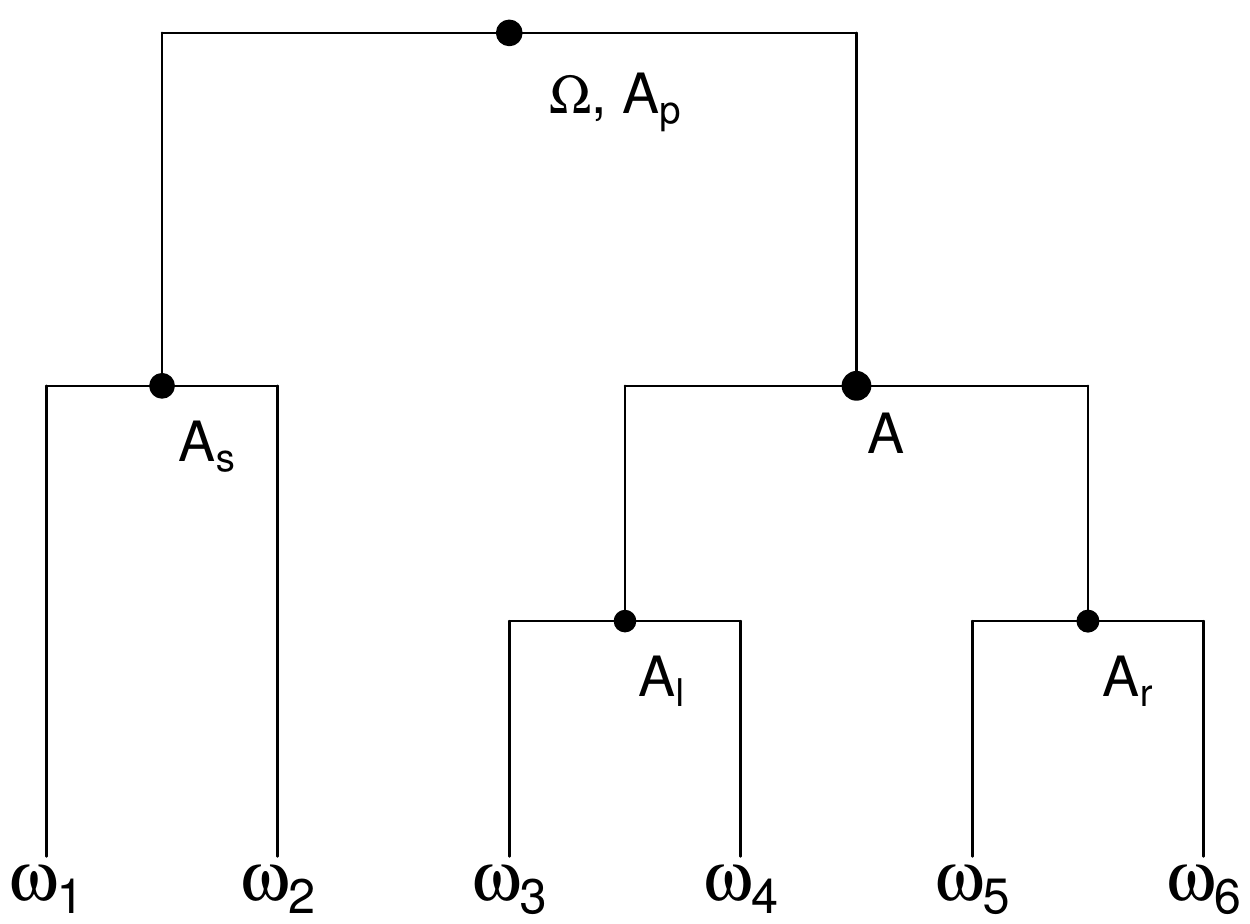}
    \captionof{figure}{{\color{black}A phylogenetic tree over $6$ OTUs.}}
        \label{fig:toy}
  \end{minipage}   
  \end{minipage} 
{\color{black}
\textit{Covariates.} Microbiome studies often collect covariate information about the samples. For example, these covariates may include the demographic information of the participants, their dietary habits, and histories of disease. For the $j$-th sample in group $i$, let $\bx_{ij}=(1,x_{ij1},\ldots,x_{ijp})'\in\mathbb{R}^{p+1}$ denote the $p$ covariates that we want to include in the analysis. Moreover, let $z_{ij}\in\{0, 1\}$ be its group indicator.}

\cite{la:2012} used the Dirichlet-multinomial model (DM) to account for the within-group heterogeneity among the OTU counts:
\begin{equation}
\begin{aligned}
\by_{ij} \mid N_{ij}, \bpi_{ij} &\overset{\mathrm{ind}}{\sim} \mathrm{Multinomial}(N_{ij}, \bpi_{ij})\\
\bpi_{ij}\mid \bpi_i, \nu &  \overset{\text{i.i.d.}}{\sim} \mathrm{Dirichlet}(\nu\bpi_i)\\
\end{aligned}
\end{equation}
for $i=0, 1$, where $\bpi_i=(\pi_{i1},\ldots, \pi_{iK})$ with $\sum^{K}_{l=1} \pi_{il}=1$, and $\nu>0$ a dispersion parameter that controls the within-group variability. A limitation of DM is its induced correlations among OTUs---that they are independent up to the summation constraint \citep{wang2017dirichlet}
, which is unrealistic since the OTUs have inherent and complicated relationships with each other, partly reflected in their evolutionary history summarized in a phylogenetic tree.

\cite{wang2017dirichlet} propose to adopt the Dirichlet-tree multinomial model (DTM) by directly incorporating the phylogenetic tree in the modeling of the OTU counts. 
Specifically, 
DTM models how the counts in $A\in\mI$ are distributed to its two children by a beta-binomial model:
\begin{equation}
\begin{aligned}
y_{ij}(A_l)\mid y_{ij}(A), \theta_{ij}(A) &\overset{\mathrm{ind}}{\sim} \mathrm{Binomial}(y_{ij}(A) , \theta_{ij}(A))\\
\theta_{ij}(A)\mid \theta_i(A), \nu(A) & \overset{\text{i.i.d.}}{\sim}\mathrm{Beta}(\theta_{i}(A)\nu(A), (1-\theta_i(A))\nu(A)),
\end{aligned}
\label{eq:bb}
\end{equation}
where $\theta_{ij}(A)$ denotes the proportion of counts in node $A$ that are distributed to its left child $A_l$, $\theta_i(A)$ the {\color{black}mean} of the $\theta_{ij}(A)$'s in group $i$, and $\nu(A)$ a precision or dispersion parameter that controls the  variability of $\theta_{ij}(A)$ {\color{black}around the group mean}, for $i=1,2$, $j=1,\ldots, n_i$. Let $\mL_{BB}(\theta_{i}(A), \nu(A) \mid y_{ij}(A_l), y_{ij}(A_r))$ denote the likelihood of the parameters in (\ref{eq:bb}) obtained by integrating out $\theta_{ij}(A)$, which with slightly more general notations can be written as 
\begin{equation}
\begin{aligned}
 \mL_{BB}(\theta,\nu\mid y_1, y_2) = 
\begin{cases}
    B(\theta\nu+y_1, (1-\theta)\nu + y_2)/B(\theta\nu, (1-\theta)\nu), & \text{if $\nu<\infty$}\\
    \theta^{y_1}(1-\theta)^{y_2}, & \text{if $\nu=\infty$}
  \end{cases}
\end{aligned}
\label{eq:bblike}
\end{equation}
for $y_1, y_2\in\{0, 1,2,\ldots,\}, \theta\in [0,1]$ and $\nu\in (0,\infty]$, where $B(\cdot,\cdot)$ is the Beta function. It can be shown that the DM likelihood for each sample can be factorized into a series of beta-binomial likelihoods along the tree  
\begin{equation}
\begin{aligned}
\mL_{DM}(\bpi_i, \nu\mid\by_{ij}) = \prod\limits_{A\in\mI} \mL_{BB}(\theta_{i}(A), \nu_i(A) \mid y_{ij}(A_l), y_{ij}(A_r))
\end{aligned}
\label{eq:l}
\end{equation}
provided that $\theta_i(A)=\pi_i(A_l)/\pi_i(A)$ and {\color{black}$\nu_i(A)=\nu\pi_i(A)$}, where $\pi_i(A)=\sum_{l\in A}\pi_{il}$ for $A\in\mI$ \citep{dennis1991hyper, dennis1996bayesian}. In this sense, DTM could be seen as a generalization of DM.


\subsection{Bayesian compositional regression for microbiome data} 

{\color{black}The likelihood factorization (\ref{eq:l}) suggests a ``divide-and-conquer'' strategy to perform inference on $\bpi_i$. Specifically, inference on $\bpi_i$ could be achieved equivalently by doing inference on $\{\theta_i(A): A\in\mI \}$, $i=0, 1$. In this section, we use this strategy to transform the original problem of comparing microbial compositions of two groups of sample into a series of tests on $\{\theta_i(A): A\in\mI \}$ in a more general situation that allows adjustment for covariates.}

For each $A\in\T$, we modify the beta-binomial specification in DTM to allow regression adjustments for covariates:
\begin{equation}
\begin{aligned}
y_{ij}(A_l)\mid y_{ij}(A), \theta_{ij}(A) &\overset{\mathrm{ind}}{\sim} \mathrm{Binomial}(y_{ij}(A) , \theta_{ij}(A))\\
\theta_{ij}(A)\mid \bx_{ij},z_{ij}, \nu(A) & \sim \mathrm{Beta}(\theta_{\bx_{ij},z_{ij}}(A)\nu(A), (1-\theta_{\bx_{ij},z_{ij}}(A))\nu(A))\\
g(\theta_{\bx_{ij},z_{ij}}(A)) & =\bx_{ij}^\top\bbet(A)+z_{ij}\gamma(A),
\end{aligned}
\label{eq:bbr}
\end{equation}
where $g: [0,1] \rightarrow \mathbb{R}$ is a link function, such as the logit link $g(x)=\log(x/(1-x))$ for $x\in (0, 1)$. $\bbet(A)\in\mathbb{R}^{p+1}$ and $\gamma(A)\in\mathbb{R}$ are the unknown parameters of the local model on $A$. {\color{black}We refer to this model as the Bayesian compositional regression (BCR).} Under BCR, testing the original null {\color{black}that there is no cross-group difference in the microbial composition} is equivalent to jointly testing a set of local hypotheses on all $A\in\mI$:
\begin{equation}
\begin{aligned}
H_{0}(A): \gamma(A) = 0 \quad \mathrm{vs} \quad H_{1}(A): \gamma(A) \not= 0. 
\end{aligned}
\label{eq:t3}
\end{equation}

Instead of modeling the OTU counts around some ``group-specific centroid'' as in DM and DTM, we model them around some ``covariate-specific centroid'' with the grouping information included as a special ``covariate''. Recent literature has shown that many covariates have associations and possible effects on the microbiome composition \citep{wang2017dirichlet, wadsworth2017integrative, xia2013logistic, tang2017mixed, ren2017bayesian, grantham2017mimixf}. {\color{black}In the testing scenario, some of these covariates may also be correlated with the grouping variable. Ignoring such confounding can reduce the power for identifying differences, or worse yet, lead to false positives. The regression adjustments in (\ref{eq:bbr}) block out the possible confounders and give us an opportunity to improve the accuracy of the testing results and reduce the chance of false discoveries.} 

{\color{black}We shall take a Bayesian approach to the local testing problem (\ref{eq:t3}) for three reasons: (1) it provides a natural way to deal with the dispersion $\nu(A)$, essentially a nuisance parameter, through integration instead of potentially unstable points estimates; (2) it allows the introduction of dependencies among the local hypotheses in a generative manner as shown in Section~\ref{subsec:graph}; and (3) it gives a fully probabilistic characterization of the uncertainty.}

To this end, we first need to specify priors for $\bbet(A)$ and for $\gamma(A)$ under $H_1(A)$. A simple choice is to put independent normal priors on the elements of $\bbet(A)$ and $\gamma(A)$. Note that using vague proper priors on the model specific parameters $\gamma(A)$ is often problematic in testing and would cause the so-called ``Bartlett's Paradox'' \citep{berger2001objective}. 
Alternatively, various principles for constructing ``objective'' priors for the coefficients can be adopted. For example, generalizations of the $g$-prior and the mixture of $g$-priors \citep{liang2008mixtures} for linear regressions to GLMs can be employed here, see for example \cite{held2015approximate} and \cite{li2015mixtures}. Specifically, one could apply the local information metric (LIM) $g$-prior as suggested by \cite{li2015mixtures} on $\gamma(A)$. In our setting, the difference in dimensions between the parameter space under the alternative and the null is only one; therefore, putting normal prior $\mathrm{N}(0, \sigma^2_{\gamma}(A))$ with reasonably large variance (such as $ \sigma^2_{\gamma}(A)=10$) would give reasonable results. For the ``common'' parameters $\bbet(A)$, we adopt independent normal priors $\mathrm{N}(0,\sigma^2_{\bbet}(A))$ with large $\sigma_{\bbet}(A)$ on its elements. For example, when the covariates are standardized (rescaled to have a mean of zero and a standard deviation of one), $\mathrm{N}(0, 16)$ covers most probable values of $\bbet(A)$ in many applications.

Without further knowledge about the nuisance dispersion parameter $\nu(A)$, we put a prior $G_A(\nu)$ on it and integrate it out. This is key to the improvement gained by applying a Bayesian testing scheme compared to its frequentist counterparts that rely on a point estimator of $\nu(A)$ (\ref{sec:num_exam}). It is necessary for $G_A(\nu)$ to have a large support to allow various levels of dispersions. {\color{black}We hence 
take $\log_{10}\nu(A) \sim \text{Unif}(-1, 4)$ that covers a wide range of dispersion levels but does not put too much prior mass on unreasonably large or small values \citep{ma:2016}}. Other choices with unbounded support such as Gamma priors on $\nu(A)$ can also be adopted.

To perform posterior inference, we introduce the state indicator of the test on node $A$ for $A\in\mI$:
\begin{equation}
S(A)=
\begin{cases}
    0, & H_0(A) \text{ is true,}\\
    1, & H_1(A) \text{ is true.}
  \end{cases}
  \label{eq:state}
\end{equation}
{\color{black}We assume for now that the $S(A)$'s are independent and let $\Pr(S(A)=1) = \rho(A)$ \textit{a priori}}. The testing problem on $A$ now becomes an inference problem on $S(A)$. Let $M_s(A)$ be the marginal likelihood under $H_s(A)$, $s\in\{0, 1\}$. We have
\begin{equation*}
\begin{aligned}
M_s(A)&=C{\color{black}\iiint}\prod\limits^{2}_{i=1}\prod\limits^{n_i}_{j=1} \mL_{BB}(g^{-1}(\bx_{ij}^\top\bbet+z_{ij}\gamma),\nu \mid y_{ij}(A_l), y_{ij}(A_r))dF_{s,A}(\bbet)dF_{s,A}(\gamma)dG_A(\nu),
\end{aligned}
\end{equation*}
where $C$ is a constant with respect to the parameters, $F_{s,A}(\bbet), F_{s,A}(\gamma)$ are the priors for $\bbet(A)$ and $\gamma(A)$ under $H_s(A)$,  $\mL_{BB}(\cdot)$ the beta-binomial marginal likelihood defined in (\ref{eq:bblike}), and $G_A(\nu)$ the prior on $\nu(A)$. {\color{black}We give details on the computational strategy for evaluating $M_s(A)$ 
 in Online Supplementary Materials~A. Given $M_s(A)$,} the posterior of $S(A)$ is
\begin{equation*}
\begin{aligned}
S(A)\mid {\color{black}\tilde\rho(A)} &\sim \mathrm{Bernoulli}(\tilde \rho(A)), \vspace{3mm}\text{ where } \vspace{4mm} \tilde\rho(A) &= \frac{\rho(A)M_1(A)}{(1-\rho(A))M_0(A) + \rho(A)M_1(A)}.
\end{aligned}
\end{equation*}
We shall refer to $\tilde\rho(A)= {\color{black}\Pr(S(A)=1\mid \bY)}$ as the \textit{posterior marginal alternative probability} (PMAP) on node $A$, {\color{black}denoted also as $\text{PMAP}(A)$}. The PMAP is large when the evidence against the local null is strong. The set of PMAPs along $\T$ can be used to test the original null and pinpoint the differences. We postpone the details on using PMAPs for making such decisions based on multiple testing considerations to Section~\ref{subsec:dec}.

{\color{black}We end this section with three further comments on our model specification in BCR. Firstly, we assume that the two groups share a common dispersion parameter $\nu(A)$ to simplify the computation. If this assumption is not likely to hold in the data, group-specific dispersions can be adopted. 
Secondly, BCR does not automatically select covariates. 
This is not restrictive in the testing scenario since our primary goal is not to identify the best predictive model for the microbiome composition or to estimate the ``effect size'' of specific covariates on the microbiome composition, but to preclude them from introducing bias to the analysis. In practice, all the suspect confounders can be included in the model even some of them may turn out to have no confounding effects provided that the number of covariates is not too large. We note that one must practice caution in adopting covariate selection in this context. Traditional statistical variable selection is based on the predictive ability of the covariates, and can lead to inappropriate elimination of confounders and result in false positives. We provide more discussion and a simple numerical illustration in Online Supplementary Materials~C. Thirdly, in the regression setup, we assume that the regression coefficients $\bbet(A)$ are the same for both groups. This ``common slope'' assumption essentially serves as an identifiability constraint, without which the null hypothesis is ill-defined---the two groups will be identical only for a specific combination of covariate values but not others, and consequently it would not make sense to test whether two groups of samples involving different covariate values are ``identical'' in microbiome composition. }



\subsection{Bayesian graphical compositional regression }
\label{subsec:graph}

{\color{black}In BCR}, the test on node $A$ is performed based only on the empirical evidence in $A$. Let $\mS=\{S(A):A\in\mI \}$, the collection of {\color{black} state indicators of all internal nodes in $\T$}. Elements in $\mS$ are independent \textit{a priori}. However, this independence assumption disregards the inherent relations of the $S(A)$'s---that they are naturally linked by the phylogenetic tree. Introducing suitable dependency structures on $\mS$ can enhance inference because (1) cross-group differences often occur along the phylogenetic tree in a clustered manner, forming chains, which is pointed out in \cite{tang:2017} and will be confirmed in our data analysis, and (2) from a statistical perspective, some local tests may involve only few counts and their power is thus limited, especially when the nodes are close to the leaves of $\T$. Introducing a dependency structure on $\mS$ allows these nodes to borrow information from each other and thus increases the power of the tests. The dependency structure we introduce should satisfy two desiderata. On the one hand, it should allow flexible information sharing among nodes that are close on the phylogenetic tree. On the other hand, it should {\color{black}be simple enough to} keep the posterior inference tractable. 

{\color{black}Recall that the node-specific beta-binomial models in BCR are naturally linked together by the phylogenetic tree, which provides a proxy to their functional relationship. This suggests that we can use the phylogenetic information when imposing dependency structures on $\S$. In particular, we consider a ``bottom-up'' auto-regressive structure such that if the signal (cross-group difference) is present at a certain node $A\in\T$, there is a chance for it to be ``carried upwards'' to $A_p$. Formally, for $A\in\mI$, we let} 
\begin{equation}
\begin{aligned}
\mathrm{logit}[\Pr(S(A)=1\mid S(A_l), S(A_r)) ]&=\alpha(A) + \tau(A)\cdot\mathbbm{1}_{[S(A_l)+S(A_r)\geq1]} \\
&\quad +\kappa(A)\cdot\mathbbm{1}_{[S(A_l)+S(A_r)=2]}, \\
\end{aligned}
\label{eq:ag}
\end{equation}  
where $\tau(A),\kappa(A)\geq 0$ together describe how likely signals at {\color{black}$A_l$ or $A_r$} are passed upwards along the tree. {\color{black}If $A_l$ or $A_r$ (or both) is a leaf node, we set $S(A_l)$ or $S(A_r)$ to be zero since there is no test performed on that node.} This model can be embedded in a family of models specified by the conditional probabilities
\begin{equation}
\begin{aligned}
\Pr(S(A)=s\mid S(A_l)=s_l,S(A_r)=s_r)=\rho_{s_ls_r,s}(A),
\end{aligned}
\label{eq:trans}
\end{equation}
where $s,s_l,s_r\in\{0,1\}$ and $\rho_{s_ls_r,s}\in [0,1]$. Specifically, the auto-regressive model forces $\rho_{01, s}(A)=\rho_{10,s}(A)$ {\color{black}since usually there is no specific prior information to differentiate between the two children of a node}. Equivalently, the conditional probability of $S(A)$ given $S(A_l),S(A_r)$ can be characterized by a transition matrix
\begin{equation*}
\begin{aligned}
\bm{\rho}(A)=
  \begin{pmatrix}
    \rho_{00,0}(A) &  \rho_{00,1}(A) \\
    \rho_{01,0}(A) &  \rho_{01,1}(A) \\
    \rho_{10,0}(A) &  \rho_{10,1}(A) \\
    \rho_{11,0}(A) &  \rho_{11,1}(A) \\
  \end{pmatrix}
\end{aligned}
\end{equation*}
where the elements in each row of $\bm{\rho}(A)$ sum up to $1$.  {\color{black}We denote $\bm{\rho}=\{\bm{\rho}(A):A\in\mI \}$ and defer the specification of elements in $\brho$ to Section \ref{subsec:dec}. The set of auto-regressive models with transition probabilities $\brho$ induces a joint distribution $F_\S$ on $\S$. Incorporating this distribution into BCR leads to a hierarchical formulation of our model for each $A\in\mI$:
\begin{equation}
\begin{aligned}
S(A)\mid \brho & \sim F_{\S}, \quad \nu(A)\mid G_A \overset{\mathrm{ind}}{\sim}  G_A\\
\bbet(A) & \overset{\mathrm{ind}}{\sim} \mathrm{N}_{p+1}(\bzero, \sigma_{\bbet}^2(A)\cdot I_{p+1})  \\
\gamma(A) &  \overset{\mathrm{ind}}{\sim}\mathbbm{1}_{[S(A)=0]}\cdot \delta_0 + \mathbbm{1}_{[S(A)=1]}\cdot \mathrm{N}(0, \sigma_{\gamma}^2(A)) \\
g(\theta_{\bx_{ij},z_{ij}}(A)) & =\bx_{ij}^\top\bbet(A)+z_{ij}\gamma(A)\\
\theta_{ij}(A)\mid \bx_{ij},z_{ij}, \nu(A) & \sim \mathrm{Beta}(\theta_{\bx_{ij},z_{ij}}(A)\nu(A), (1-\theta_{\bx_{ij},z_{ij}}(A))\nu(A))\\
y_{ij}(A_l)\mid y_{ij}(A), \theta_{ij}(A) &\overset{\mathrm{ind}}{\sim} \mathrm{Binomial}(y_{ij}(A) , \theta_{ij}(A)),\quad i=1, 2, j =1,\ldots, n_i,
\end{aligned}
\label{eq:hierarchical}
\end{equation}
where $I_{p+1}$ is the $(p+1)$-dimensional identity matrix, $\mathbbm{1}_{[\cdot]}$ the indicator function, and $\delta_0$ a point mass at zero. We shall refer to this model as the Bayesian graphical compositional regression (BGCR).}

{\color{black}An alternative way to introduce dependencies among elements in $\S$ that also incorporate the phylogenetic information is by utilizing the ``top-town'' Markov tree model (MT) \citep{crouse1998markov, soriano2017probabilistic}, which instead specifies $\Pr(S(A)\mid S(A_p))$ for $A\in\T\setminus\{\Omega\}$.} However, the ``explaining away'' effect \citep{wellman1993explaining} of the auto-regressive model is important for microbiome data. MT pushes signals downwards along the tree to both children of a node and implies cooccurrence of signals in the sibling nodes, which is not the typical situation in the microbiome context. Instead, the signals in microbiome studies often form chains, with only one of the two children nodes share the signal with the parent node. This is the primary reason for adopting a ``bottom-up'' autoregressive model. 



\subsection{Inference under BGCR }
\label{subsec:inference}

{\color{black}
As in BCR, the marginal posteriors $\{\Pr(S(A)= 1\mid\bY): A\in\mI\}$ play a pivotal role in the posterior inference under BGCR. We next show that these quantities can be calculated exactly up to the approximation to $M_s(A)$'s, without entailing Monte Carlo simulations.

The joint distribution introduced by BGCR on $\S$ can be represented by a Bayesian network, i.e., a directed acyclic graph (DAG) $\mathcal{G}(V_{\S}, B_{\S})$ with nodes $V_\S=\{S(A):A\in\mI \}$ and edges $B_\S=\{S(A)\rightarrow S(A_p):A\in\mI\setminus\{\Omega\} \}$. Moralizing $\mathcal{G}(V_{\S}, B_{\S})$ yields an undirected graph $\mathcal{U}(V_{\S}, E_{\S})$ from which the conditional dependencies in $\S$ can be read off directly \citep{koller2009probabilistic}. Specifically, the vertices of $\U$ are the same of those of $\G$; there is an undirected edge between $S(A)$ and $S(A^\prime)$ in $\U$ if $A$ and $A^\prime\in\mI$ are siblings or if they form a parent-child pair: $E_\S = \left\{ \{S(A), S(A^\prime)\}: A^\prime = A_p \text{ or } A^\prime = A_s, A\in\mI\setminus\{\Omega \}\right\}$. 
Moreover, we refer to a chain of edges as a ``path''. Two nodes $S(A)$ and $S(A^\prime)$ in $\U$ are conditionally independent given a set of nodes $v=\{S(A_1),\ldots, S(A_m)\}\subset V_\S $ if and only if any path connecting $S(A)$ and $S(A^\prime)$ passes some elements in $v$. \ref{fig:treeillustration} (i), (ii), and (iii) respectively give illustrations of a fictional phylogenetic tree, the corresponding Bayesian network $\G$ on $\S$, and the moralized undirected graph $\U$. The structure of $\U$ induced by BGCR has two implications on posterior computations. On the one hand, $\U$ contains many loops, {\color{black}which at first glance would substantially complicate the posterior and the inference algorithm.} If $\U$ were a tree-type graph without loops, as in MT, simple and exact inference can be achieved by a forward-backward algorithm. On the other hand, all the loops in $\U$ include only three nodes and the dependencies among $S(A)$'s are local---$S(A)$ is independent of other elements in $\S$ once conditioning on $S(A_p), S(A_s), S(A_l)$ and $S(A_r)$, suggesting that $\U$ is not ``too far'' from a tree-type graph that allows exact inference. Intuitively, if we treat each loop as a unit, $\U$ can be viewed as a ``tree of loops'' on which exact inference can now be performed.

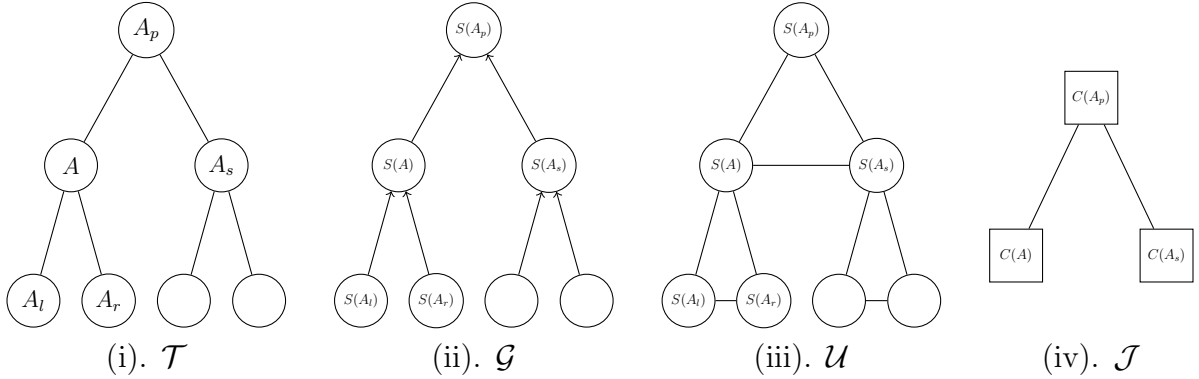
\begin{figure}[!h]
\centering
\begin{tikzpicture}
  [scale=.6,auto=left,every node/.style={circle,minimum size=20pt,inner sep=1pt}]
  \tikzset{edge/.style = {-}}

 \node (n1)[draw] at (8.75/3,12) { \scalebox{0.75}{ $A_p$ } };
 \node (n2)[draw] at (3.75/3,9)   { \scalebox{0.75}{ $A$ }  };
\node (n3)[draw] at (13.75/3,9)  {\scalebox{0.75}{ $A_s$ }  };
\node (n4)[draw] at (1.25/3,6) {  \scalebox{0.75}{ $A_l$ }  };
\node (n5)[draw] at (6.25/3,6)  { \scalebox{0.75}{ $A_r$ }  };
\node (n6)[draw] at (11.25/3,6) { };
\node (n7) [draw]at (16.25/3,6) { };
\node (n8) [style={rectangle}]  at (8.75/3, 4.7){(i). $\T$};

\draw[edge] (n1) to (n2);
\draw[edge] (n1) to (n3);
\draw[edge] (n2) to (n4);
\draw[edge] (n2) to (n5);
\draw[edge] (n3) to (n6);
\draw[edge] (n3) to (n7);
\end{tikzpicture}\hspace{0.5cm}
\begin{tikzpicture}
  [scale=.6,auto=left,every node/.style={circle,minimum size=20pt,inner sep=1pt}]
  \tikzset{edge/.style = {<-}}
 \node (n1)[draw] at (8.75/3,12) { \scalebox{0.45}{ $S(A_p)$ } };
 \node (n2)[draw] at (3.75/3,9)   { \scalebox{0.45}{ $S(A)$ }  };
\node (n3)[draw] at (13.75/3,9)  { \scalebox{0.45}{ $S(A_s)$ }  };
\node (n4) [draw]at (1.25/3,6) {  \scalebox{0.45}{ $S(A_l)$ }  };
\node (n5) [draw]at (6.25/3,6)  { \scalebox{0.45}{ $S(A_r)$ }  };
\node (n6) [draw]at (11.25/3,6) { };
\node (n7) [draw]at (16.25/3,6) { };
\node (n8) [style={rectangle}]  at (8.75/3, 4.7){(ii). $\G$ };

\draw[edge] (n1) to (n2);
\draw[edge] (n1) to (n3);
\draw[edge] (n2) to (n4);
\draw[edge] (n2) to (n5);
\draw[edge] (n3) to (n6);
\draw[edge] (n3) to (n7);
\end{tikzpicture}\hspace{0.5cm}
\begin{tikzpicture}
  [scale=.6,auto=left,every node/.style={circle,minimum size=20pt,inner sep=1pt}]
  \tikzset{edge/.style = {-}}
 \node (n1)[draw] at (8.75/3,12) { \scalebox{0.45}{ $S(A_p)$ } };
 \node (n2)[draw] at (3.75/3,9)   { \scalebox{0.45}{ $S(A)$ }  };
\node (n3) [draw]at (13.75/3,9)  {\scalebox{0.45}{ $S(A_s)$ }  };
\node (n4) [draw]at (1.25/3,6) {  \scalebox{0.45}{ $S(A_l)$ }  };
\node (n5)[draw] at (6.25/3,6)  { \scalebox{0.45}{ $S(A_r)$ }  };
\node (n6)[draw] at (11.25/3,6) { };
\node (n7) [draw]at (16.25/3,6) { };
\node (n8) [style={rectangle}]  at (8.75/3, 4.7 ){(iii). $\U$ };

\draw[edge] (n1) to (n2);
\draw[edge] (n1) to (n3);
\draw[edge] (n2) to (n4);
\draw[edge] (n2) to (n5);
\draw[edge] (n3) to (n6);
\draw[edge] (n3) to (n7);

\draw[edge] (n2) to (n3);
\draw[edge] (n4) to (n5);
\draw[edge] (n6) to (n7);
\end{tikzpicture}\hspace{0.5cm}
\begin{tikzpicture}
  [scale=.6,auto=left,every node/.style={rectangle,minimum size=20pt,inner sep=1pt}]
  \tikzset{edge/.style = {-}}
  
\node (n1)[draw] at (8.75/3,10.5) { \scalebox{0.45}{ $C(A_p)$ } };
\node (n2)[draw] at (3.75/3,7)   { \scalebox{0.45}{ $C(A)$ }  };
\node (n3) [draw]at (13.75/3,7)  {\scalebox{0.45}{ $C(A_s)$ }};  
\node (n8) [style={rectangle}]  at (8.75/3, 4.7 ){(iv). $\J$ };
  
\draw[edge] (n1) to (n2);
\draw[edge] (n1) to (n3);
\end{tikzpicture}

\caption{(i). A fictional phylogenetic tree (the leaves of $\T$ are not shown); (ii). The ``bottom-up'' Bayesian network $\G$ on $\S$ corresponds to $\T$; (iii). The undirected graph after moralizing $\G$; (iv). The clique tree derived from $\U$.}
\label{fig:treeillustration}
\end{figure}

Formally, for $S(A)\in V_\S$, we call $C(A) = (S(A), S(A_l), S(A_r) )$ the \textit{clique} of $S(A)$, or simply the \textit{clique} of $A$. The conditional dependencies among $\C=\{C(A):A\in\mI \}$ are inherited from $\U$ and can be represented by an undirected graph $\J(V_\C, E_\C)$ in a similar manner, where $V_\C = \C$, $E_\C = \{\{C(A),C(A^\prime)\} : A^\prime = A_p, A\in\mI\setminus\{\Omega \}\}$. See \ref{fig:treeillustration} (iv) for an illustration. It is easy to see that $\J$ contains no loop and that any two nodes in $\J$ are connected by a unique path. Therefore, $\J$ is an undirected tree. We refer to $\J$ as the \textit{clique tree} and perform inference thereon. Note that $C(A)$ can take on eight values, denoted by $\bc=\{c_1,c_2,\ldots,c_8\}$ where $c_i = (c_{i1},c_{i2},c_{i3})=  (\lfloor\frac{i-1}{4}\rfloor \text{ mod }2, \lfloor\frac{i-1}{2}\rfloor \text{ mod }2 , (i-1) \text{ mod }2)$ (each element is the corresponding digit of the binary representation of $i$).

For $A\in\mI$, without loss of generality, let $A$ be the left child of $A_p$. Eq.~(\ref{eq:trans}) induces the top-down transition probabilities on the clique tree $\J$:
\begin{equation}
\begin{aligned}
\Pr(C(A) = c_{i^\prime} \mid C(A_p)=c_i) &=\Pr( (S(A),S(A_l),S(A_r)) = c_{i^\prime} \mid (S(A_p),S(A),S(A_s)) = c_i)\\
& = \Pr( (S(A_l),S(A_r)) = (c_{i^\prime 2},c_{i^\prime 3})\mid S(A) = c_i),
\end{aligned}
\label{eq:tran2}
\end{equation}
for $C(A)\in \C$, where $i,i^\prime=1,2,\ldots, 8$. As with $\brho(A)$, these transition probabilities could be organized into an $8 \times 8$ matrix $\bxi(A)$ with the $i^\prime$-th element in row $i$ defined by $\bxi_{ii^\prime}(A) = \Pr(C(A) = c_{i^\prime} \mid C(A_p)=c_{i})$. Each row of this matrix represents the conditional distribution of $C(A)$ when $C(A_p)$ takes one of the eight possible values in $\bc$. Since $\Omega\in\T$ does not have a parent, we let each row of $\bxi(\Omega)$ be the induced marginals of $C(\Omega)$.

}

In addition, for $A\in\mI$, let $\T(A)$ be the subtree of $\T$ with $A$ as the root. Recall that for the $j$-th sample in group $i$, $y_{ij}(A)$ denotes the OTU counts in node $A$. Let $\by_{ij}(A)$ denote all the OTU counts that fall into the subtree $\T(A)$. Note that the difference between $y_{ij}(A)$ and $\by_{ij}(A)$ is important: $y_{ij}(A)$ contains only the counts in $A$ while $\by_{ij}(A)$ contains {\color{black}the set of $y_{ij}(A_d)$'s} for all $A_d\in\T(\A)$. Let $\bY(A)=\{ \by_{ij}(A):i=1,2, j=1,\ldots, n_i \}$ be the counts in $\T(A)$ from all the samples. Bayesian inference on $S(A)$ relies on evaluating the posterior transition probability matrix $\tilde\bxi(A)$ for each $C(A)\in \C$. We next describe how these posteriors can be calculated with a recursive algorithm. Specifically, the root of the clique tree, $C(\Omega)$, first collect information iteratively from other nodes. After all the information is collected, it is distributed downwards along the tree to finish the information sharing.

\textit{Information collection}. {\color{black}For $i=1,2,\ldots, 8$, let $\bphi_{i}(A) = \Pr(\bY(A) \mid C(A_p)=c_i)$ be the prior predictive distribution of $\bY(A)$ evaluated at the observed counts given $C(A_p)=c_i$. When $A\in \mI\setminus \{ \Omega\}$, it is given by
\begin{equation}
\begin{aligned}
\bphi_{i}(A) & := \Pr(\bY(A)\mid C(A_p)=c_i)\\
&=\sum\limits_{1\leq i^\prime\leq 8} \Pr(\bY(A)\mid C(A)=c_{i^\prime}) \Pr(C(A)=c_{i^\prime}\mid C(A_p)=c_i)\\ 
&= \sum\limits_{1\leq i^\prime\leq 8} \bxi_{ii^\prime}(A) M_{c_{i^\prime 1}}(A) \Pr(\bY(A_l)\mid C(A)=c_{i^\prime})\Pr(\bY(A_r)\mid C(A)=c_{i^\prime}) \\
&= \sum\limits_{1\leq i^\prime\leq 8} \bxi_{ii^\prime}(A)  M_{c_{i^\prime 1}}(A)  \bphi_{i^\prime}(A_l)\bphi_{i^\prime}(A_r).
\end{aligned}
\label{eq:collect}
\end{equation}
When $A$ has no children in $\mI$, by definition 
$\bphi_{i}(A)=1$ for all $i$. 
When $A=\Omega$, $C(A)$ does not have a parent in the clique tree. To simplify the notation
, we can introduce an imaginary 
node $\Omega_p$ that serves as $\Omega$'s parent. Without loss of generality, we let $\Omega$ be the left child of $\Omega_p$ and let $\Omega_s$ with $\bphi_{i}(\Omega_s) = 1$ be $\Omega$'s ``imaginary'' sibling. By setting $\Pr(C(\Omega)=c_{i^\prime}\mid C(\Omega_p)=c_i)$ to the corresponding marginals of $C(\Omega)$ for $c_i, c_{i^\prime}\in\bc$, we can keep the formulation in (\ref{eq:collect}).

As a byproduct of the inference algorithm, $\phi_1(\Omega)$ gives the marginal likelihood of the data, which can be used to guide the selection of hyper-parameters (Section~\ref{subsec:hyper}). Moreover, since the marginal likelihood is available, variable selection can be achieved by putting spike-and-slab priors on the regression coefficients (details are in Online Supplementary Materials~C). 
 }

\textit{Information distribution}. By Bayes' theorem, the posterior transition probability {\color{black}
\begin{equation}
\begin{aligned}
\tilde\bxi_{ii^\prime}(A) & :=\Pr(C(A)=c_{i^\prime} \mid C(A_p)=c_i, \bY ) \\
&=\frac{\Pr(C(A)=c_{i^\prime}, \bY(\Omega) \mid C(A_p)=c_i)}{\Pr(\bY(\Omega) \mid C(A_p)=c_i)} \\
&= \frac{\Pr(C(A)=c_i{^\prime} \mid C(A_p)=c_i)\Pr(\bY(A)\mid C(A)=c_i{^\prime})}{\Pr(\bY(A) \mid C(A_p)=c_i)} \\
&= \frac{\bxi_{ii^\prime}(A)M_{c_{i^\prime 1}}(A)\bphi_{i^\prime}(A_l)\bphi_{i^\prime}(A_r)}{\bphi_{i}(A)}.
\end{aligned}
\label{eq:distribute}
\end{equation}
}

Let $\tilde\bxi(A)$ be the matrix of the posterior transition probabilities of $C(A)$ given $C(A_p)$. Based on $\{\tilde\bxi(A): A\in \mI \}$, it is easy to compute the PMAP on each node $A\in\mI$. Specifically, starting from the root of the tree, each row of $\tilde\bxi(\Omega)$ represents the {\color{black}posterior} marginals of the clique $C(\Omega)$. Thus
\begin{equation}
\begin{aligned}
\Pr(C(\Omega)=c_{i^\prime} \mid \bY)= \sum_{1\leq i\leq 8}\tilde\bxi_{ii^\prime}(\Omega)/8
\end{aligned}
\label{eq:distribute}
\end{equation}
for $i^\prime = 1,2,\ldots,8$. By marginalization, we can get {\color{black}$\text{PMAP}(\Omega)$}. The PMAPs on other nodes can be computed by induction. For example, given the clique marginals of $C(A_p)$, we can compute the clique marginals of $C(A)$
\begin{equation}
\begin{aligned}
\Pr(C(A)=c_{i^\prime}\mid\bY )= \sum_{1\leq i\leq 8 }\Pr(C(A_p)=c_i\mid\bY)\tilde\bxi_{ii^\prime}(A) 
\end{aligned}
\label{eq:distribute}
\end{equation}
for $i^\prime = 1,2,\ldots,8$, from which the PMAP on $A$ can be obtained by marginalization. We can also compute the \textit{posterior joint alternative probability} (PJAP) that captures the empirical evidence against the global null. Formally, 
\begin{equation}
\begin{aligned}
\text{PJAP}=1-\Pr(S(A)=0, A\in\T\mid\bY)=1-\prod\limits_{A\in \mI}\tilde\bxi_{11}(A).
\end{aligned}
\end{equation}
Note that the above steps are a variant of the junction tree algorithm \citep{lauritzen1988local} that efficiently calculate the marginals in a graphical model. Directly applying the standard junction tree algorithm on $\J$ outputs the PMAPs, however, it does not allow efficient computation of the PJAP. Algorithm \ref{alg:BGCR} summarizes the entire inference recipe for testing the existence of cross-sample differences under BGCR.

\begin{algorithm}[p]
\caption{BGCR for comparing microbiome composition}
\label{alg:BGCR}

\begin{algorithmic}
\\

{\color{black}Construct the clique tree $\J$.} \Comment{Preprocessing} \\
\vspace{-0.5em}
\For{$A$ in $\mI$}   
\State{Compute the marginal likelihoods of the local test on $A$.}
\State{Compute the prior transition matrix $\bxi(A)$ on the clique of $A$.}
\State{Compute $D(A)$ --- the depth of $A$ defined as the number of edges from $A$ to the root of $\T$.}
\EndFor\\


\For{$d$ in $\max\limits_{A\in\mI}\{ D(A)\} : 0$}  \Comment{Recursive information collection}\\
 \vspace{-0.5em}
\For{$A$ with $D(A)=d$}  
 
\If{$C(A)$ has no children in $\J$}
\State{Let $\bphi_{i }(A_l)=\bphi_{i }(A_r)=1$, $i  = 1,2,\ldots, 8$.}
\Else 
\State{\text{Compute}
$ \bphi_{i}(A) = \sum\limits_{1\leq i^\prime \leq 8} \bxi_{ii^\prime}(A)  M_{c_{i^\prime 1}}(A)  \bphi_{i^\prime}(A_l)\bphi_{i^\prime}(A_r), \quad i = 1,2,\ldots, 8.$  }

\EndIf  
\EndFor
\EndFor\\

\For{$A$ in $\mI$}  \Comment{Information distribution}\\

\vspace{-0.5em}
\begin{equation*}
\begin{aligned}
\text{Compute}\quad \tilde\bxi_{ii^\prime}(A) =\frac{\bxi_{ii^\prime}(A)M_{c_{i^\prime 1}}(A)\bphi_{i^\prime}(A_l)\bphi_{i^\prime}(A_r)}{\bphi_{i}(A)},\quad i,i^\prime = 1,2,\ldots, 8.
\end{aligned}
\label{eq:distribute}
\end{equation*}

\EndFor\\
  \\
{\color{black}Compute the PJAP.   \Comment{Global information summary}}\\

\For{$d$ in $0:\max\limits_{A\in\mI}\{ D(A)\} $} \Comment{Local information summary}\\
\vspace{-0.5em}
\For{$A$ with $D(A)=d$}  
\If{$A=\Omega$}
\State{Let $\Pr(C(A_p)=c_i \mid\bY)=1/8$,\quad i = 1,2,\ldots, 8.}
 \Else

\State{\text{ Compute} $\Pr(C(A)=c_{i^\prime} \mid\bY) = \sum_{1\leq i\leq 8}\Pr(C(A_p)=c_i\mid\bY)\tilde\bxi_{ii^\prime}(A)$, $i^\prime=1,2,\ldots, 8$. }

\State{\text{ Compute PMAP$(A)$ by marginalization.} }

\EndIf
\EndFor
\EndFor\\
 
\State{Report nodes with $\text{PMAP}(A) > L$ for some threshold $L$.} \Comment{Decision making}
\vspace{0.5mm}
{\color{black}\State{Reject the original null if $\text{PJAP} > 0.5$.}  }
\end{algorithmic}
\end{algorithm}

 
\subsection{Decision making}
\label{subsec:dec}

{\color{black}The PJAP and the PMAPs provide the bases of making decisions about the original and the node-specific hypotheses along the tree. For the original testing problem, we reject the null if $\text{PJAP} > 0.5$, which corresponds to the Bayes optimal decision rule under the simple 0-1 loss. Similarly, we reject $H_0(A)$ and call $A$ a significant node if $\text{PMAP}(A) > L$ for some threshold $0<L<1$. For example, $L=0.5$ is recommended by \cite{barbieri2004optimal}. Details on decision making can be found in Online Supplementary Materials~B. It is worth noting that by reporting the significant nodes and marking them on the phylogenetic tree (or simply mark the PMAPs along the tree without explicitly providing a decision), we have a natural way to characterize and visualize the cross-group differences that offers more insights than merely providing a decision about a test on the original null. For example, it sheds light on the set of microbes most relevant to the cross-group differences.}


\subsection{Specifications of hyper-parameters}
\label{subsec:hyper}
In BGCR, we need to specify 
$\alpha(A),\tau(A)$ and $\kappa(A)$ in (\ref{eq:ag}). For simplicity, we let $\alpha, \tau, \kappa$ be global parameters that are the same for all $A\in\mI.$ 

\textit{Choice of $\alpha$}. The \textit{prior joint alternative probability} (PrJAP) can be written as 
 \begin{equation*}
 \begin{aligned}
\mathrm{PrJAP}=1-\left( \frac{1}{1+\exp(\alpha)} \right)^{|\mI|}, \\
\end{aligned}
\end{equation*} 
 which is a monotone function of $\alpha$. We can then choose $\alpha=\alpha_0$ such that the PrJAP is at a desired level such as $0.5$.
 
 \textit{Choice of $\kappa$}.  $\tau$ and $\kappa$ together control the ``stickiness'' of the signals, that is, {\color{black}how likely the ``chaining'' pattern will occur} along the phylogenetic tree. 
 To encapsulate the ``explaining away'' effect, we set $\kappa=0$ and assume that the signal at $A$ could be ``explained away'' by the signal at one of its children. Another choice for $\kappa$ is $\kappa=\tau$, which imposes additive effect of signals at the sibling nodes on the parent. 
 
 \textit{Choice of $\tau$}.  Given $\alpha=\alpha_0, \kappa = 0$, we use an empirical Bayes procedure to choose the value of $\tau$. Given $\tau$, the marginal likelihood is given by $\bphi_{1}(\Omega)$ as defined in (\ref{eq:collect}). We maximize $\bphi_{1}(\Omega)$ over $\tau$ to get $\hat\tau$ as the prior choice for $\tau$. In search of $\tau$, we focus on the interval $[0, \tau_{max})$, where $\tau_{max}$ is some predetermined upper bound on $\tau$. One way to choose $\tau_{max}$ is based on the \textit{prior marginal alternative probability} (PrMAP) on each node, which is a monotonically increasing function of $\tau$. Therefore, the total number of nodes with PrMAP exceeding the threshold $L$ also increases with $\tau$. We can preselect an upper bound of the total prior expected number of true alternatives and solve for $\tau_{max}$ accordingly. In practice, when the number of OTUs is not very large ($\leq$ 150), we find that setting $\tau_{max}=6$ usually works well (this corresponds to setting the prior expected sum of PrMAPs equal to $2$ with 100 total OTUs). When $\tau=0$, BGCR incorporates BCR as a special case.


\section{Numerical examples}
\label{sec:num_exam}


\subsection{Evaluating the performance of BGCR}

We first carry out three simulation studies to illustrate the performance of BGCR. {\color{black}For these simulations, several synthetic datasets are obtained} based on the July 29, 2016 version of the fecal data of the American Gut Project \citep{mcdonald2015towards}, which collected microbiome samples from different body sites of a large number of participants and offers publicly accessible datasets that can be downloaded from \url{http://americangut.org}. {\color{black}The full dataset we use contains counts of 27774 OTUs from 8327 fecal samples. Each sample is taken from a unique participant individual; $455$ covariates describing various aspects of the subjects are collected, such as their demographical information and dietary behaviors. Moreover, a {\color{black} rooted} full binary phylogenetic tree on all OTUs is available.}

Although the number of OTUs in the study is large, many cells in the OTU table contain zero {\color{black}or very few} counts. For illustration purpose, we focus on the top $50$, $75$, and $100$ OTUs with the largest overall counts across samples. {\color{black}Using different numbers of OTUs allow us to evaluate the inference of increasing dimensionality on the performance of different tests. Note that even with these filtered top OTUs, the data still demonstrate the four challenging features of the OTU counts listed in \ref{sec:intro}.} We further narrow down the samples to $561$ middle-aged (people in their 30s, 40s and 50s) male Caucasian participants from the west census region to reduce the large variation across different simulation rounds (we will use the full dataset in our later data analysis). Reducing the {\color{black}sample size in} the simulation is also necessary due to the speed of the competitors. In each simulation, we randomly divide the data into two {\color{black}roughly equal-sized groups to create} the data under the null. The data under the alternative is generated under three scenarios:
 
\begin{enumerate}
\item [$\RN{1}$.] {\textbf{Cross-group difference exists at a single OTU.}}
This scenario is also considered in \cite{tang:2017}. In each round of the simulation, we randomly select an OTU and increase its count in the second group by a given percentage $p$. {\color{black}This induces cross-group differences on the parent node of the selected OTU.} In this case, the differences are considered local on the tree. We consider $K=50, 75$, and $100$ with $p = 100\%, 150\%$, and $250\%$ (To rule out obvious simulations with either too weak or too strong signals, we place additional constraints in the random selection. Specifically, we only select OTUs with sample means in the middle $80\%$ range of all the $K$ OTU sample means).

\item [$\RN{2}$.] {\textbf{Cross-group difference exists at multiple OTUs.}}
Similar to the first scenario, $eight$ OTUs are randomly selected and their counts are increased by a given percentage $p$ of the subjects in the second group to create cross-group differences that are more global on the tree. We consider $K=50, 75$, and $100$ with $p = 50\%, 100\%$, and $150\%$.

\item [$\RN{3}$.] {\textbf{Cross-group difference exists at a chain of nodes in the phylogenetic tree.}}
Cross-group differences of OTU compositions often cluster into chains as shown in \ref{sec:app} and \cite{tang:2017}. {\color{black}To see how the graphical structure in BGCR helps increase the power of the test in this situation, we consider a case in which cross-group differences are present in} a fixed chain of nodes in the phylogenetic tree. {\color{black}In particular, we focus on the top $100$ OTUs ($K=100$) with their phylogenetic tree shown in \ref{fig:sim3_tree}. Consider a chain of three nodes in the tree as shown in red in \ref{fig:sim3_tree}, we create cross-group differences at these nodes by systematically modifying the counts of their descendant OTUs.} Specifically, we increase the counts of OTU 1, 2 and 3 (marked in \ref{fig:sim3_tree}) of all the subjects in the second group by $0.33 p, 0.67 p$ and $p$ percent, respectively. We consider the case when $p =75\%, 100\%$, and $125\%.$
\begin{figure}[!ht]
\begin{center}
\includegraphics[width = 12cm]{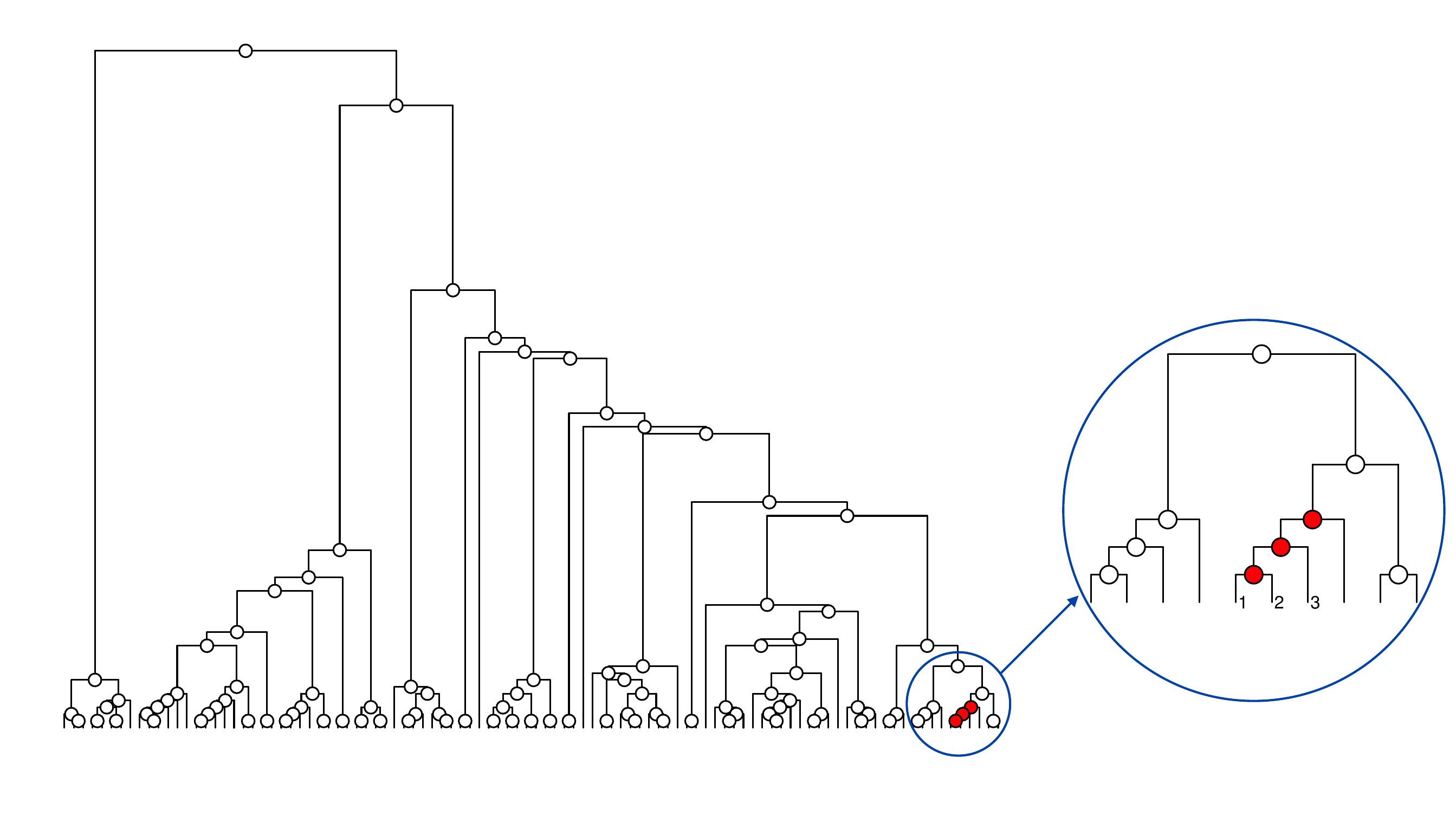}\\
\caption{{\color{black}The phylogenetic tree of the top $100$ OTUs with the largest overall counts. Each circle represents an internal node; each tip represents an OTU. The chain of nodes with designed cross-group differences in simulation $\RN{3}$ is marked red. The zoom-in plot gives a better view of these nodes and their descendant OTUs.}}
\label{fig:sim3_tree}
\end{center}
\end{figure}
\end{enumerate}

In each scenario, we carry out 3000 rounds of simulations. For BGCR, we let $\gamma(A)\sim \rm{N}(0, 10)$ and $\log_{10}\nu(A)\sim \text{Unif}(-1, 4)$ independently for $A\in\mI$. Hyperparameters are chosen according to Section \ref{subsec:hyper} with $\text{PrJAP}=0.5$, $\kappa=0$, and $\tau_{max}=6$. We compare BGCR with DTM using the maximum of the single node statistic (DTM-1), using the maximum of the triplet statistic (DTM-3) \citep{tang:2017} and the DM test \citep{la:2012}.

\begin{figure}[!ht]
\begin{center}
\includegraphics[width = 16.5cm]{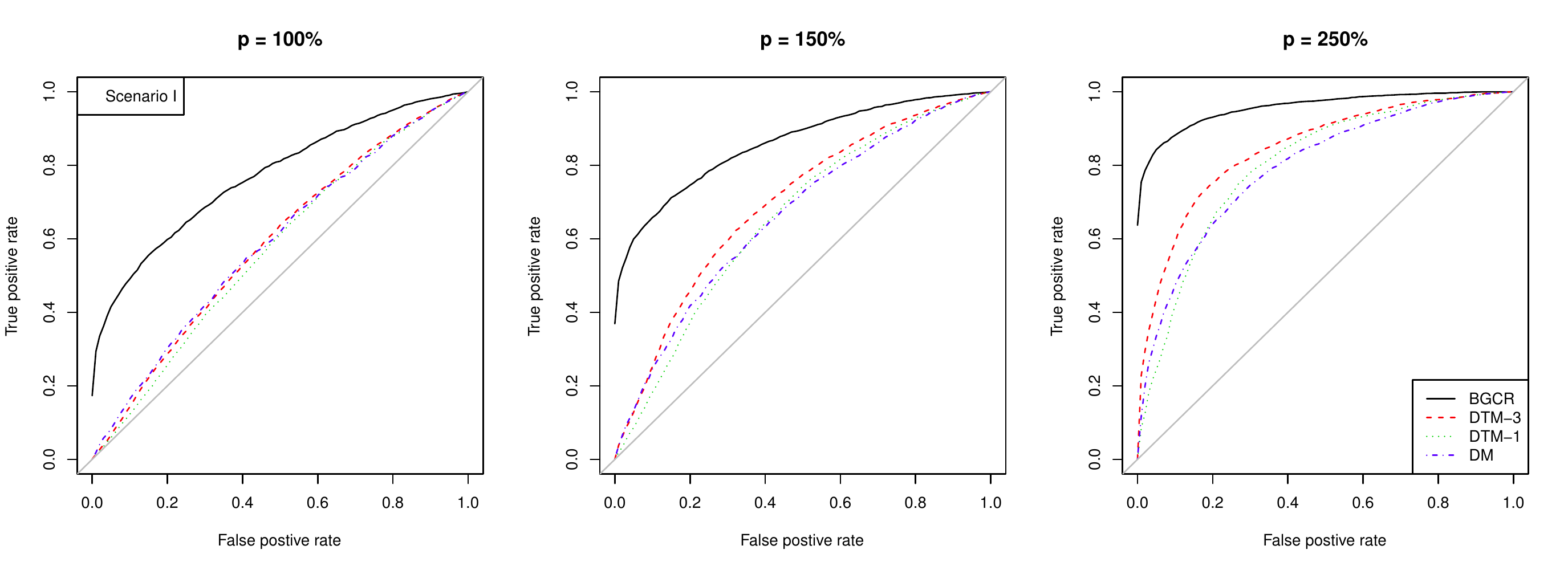}\\
\includegraphics[width = 16.5cm]{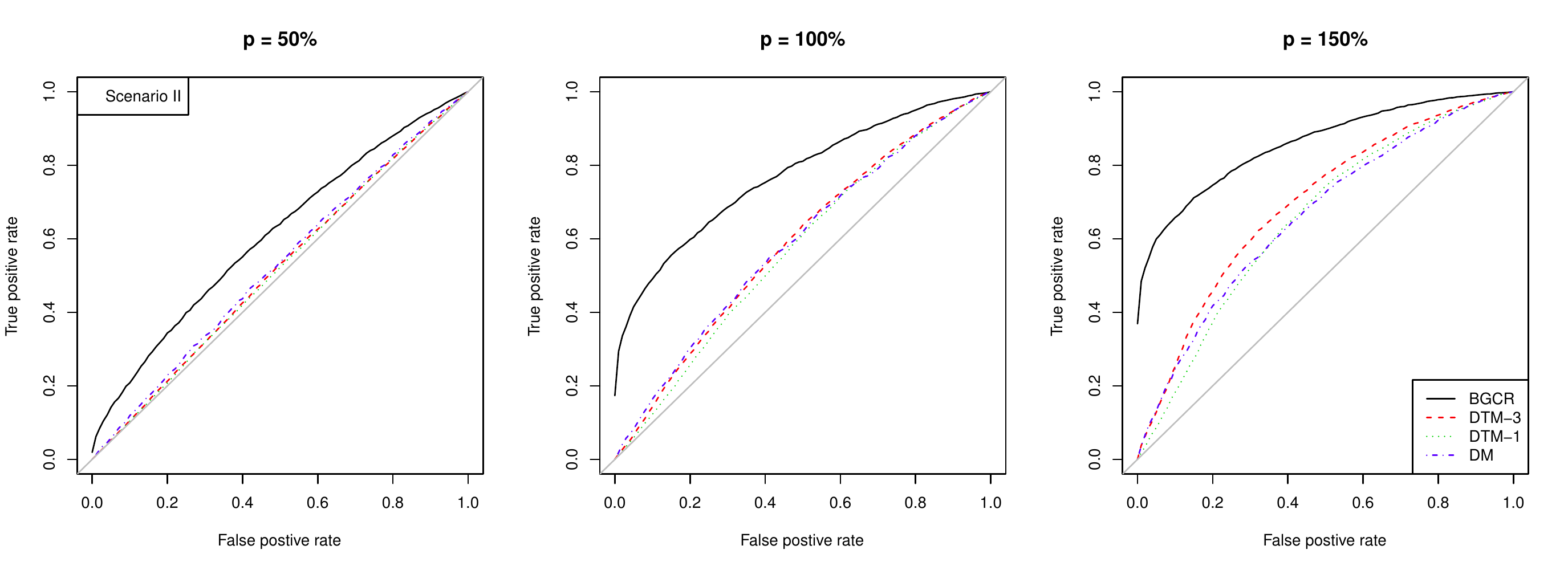}\\
\includegraphics[width = 16.5cm]{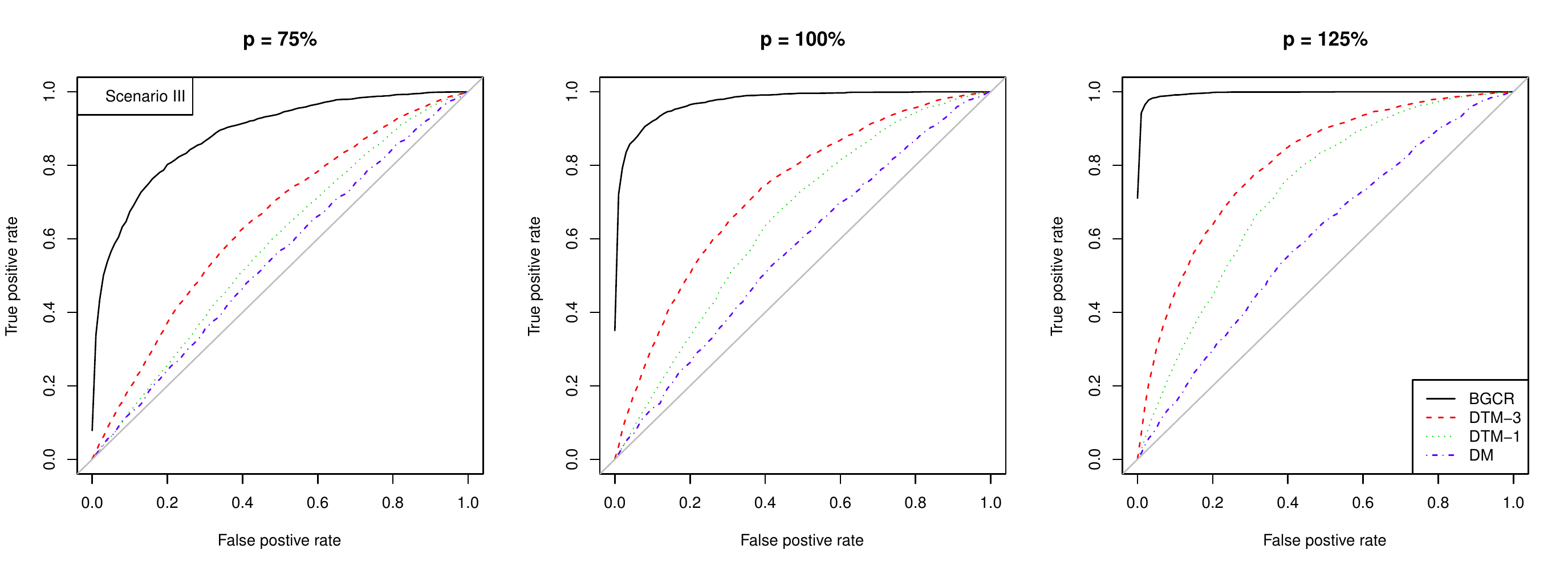}\\
\caption{ROC curves for Scenario $\RN{1}$ (top row), $\RN{2}$ (middle row) and $\RN{3}$ (bottom row) with $K = 100$. The columns are indicated by the percent of counts increased in the second group ($p$). }
\label{fig:roc_1}
\end{center}
\end{figure}

The ROC plots under {\color{black}the three} simulation scenarios with $K =100$ are shown in \ref{fig:roc_1}. Similar results under Scenario $\RN{2}$ and $\RN{3}$ with $K=50$ and $K=75$ and shown in Figure S1 and Figure S2 in online supplementary materials. Overall, BGCR outperforms the competitors in all these scenarios. In general, the three methods that incorporate phylogenetic information perform better than the DM test. In Scenario $\RN{1}$ and $\RN{2}$, \ref{fig:roc_1} and Figure S1 show that DTM-1 and DTM-3 work better when $K$ is small. {\color{black}Compared with BGCR, these tests 
perform worse with the increase of $K$.} This is due to the inherent features of the microbiome data. When $K$ is small, the samples typically contain large counts of {\color{black}many} OTUs included in the study and are likely to be described by the DM-type models. When $K$ becomes larger, the OTU table gets sparser, making it harder for DTM-1 and DTM-3 to estimate the right dispersion parameters. In contrast, the dispersion parameters are integrated out rather than estimated in BGCR.


To illustrate how borrowing information among neighboring nodes helps increase the power of the node-specific tests, we compare BGCR with BCR in Scenario $\RN{3}$. Under the null, the estimated $\gamma$ by the empirical Bayes procedure in BGCR are zero in all but a few of the simulations (Figure S3 left). Therefore, BGCR essentially degenerates to BCR and does not increase the chance of false positives (Figure S3 middle and right). Under the alternative, by introducing positive dependencies among the local tests, BGCR obtains larger PJAPs and thus provides a better chance to correctly reject the null. To see how this would influence the decision procedure, we show the proportion of tests that reject the null under different decision thresholds $L$ in Figure S4 in online supplementary materials. When $0.5<L<0.9$, the gain by introducing the dependencies among tests is remarkable. 
Figure S5 in online supplementary materials shows the estimated $\gamma$'s under the alternatives. When the cross-group difference increases, the chaining pattern gets stronger, causing the estimated $\gamma$ to increase. Figure S6 in online supplementary materials shows the corresponding PJAPs of the two models under comparison.

 \subsection{Visualizing the cross-group differences with BGCR}

\begin{figure}[!h]
\begin{center}
\includegraphics[width = 15cm]{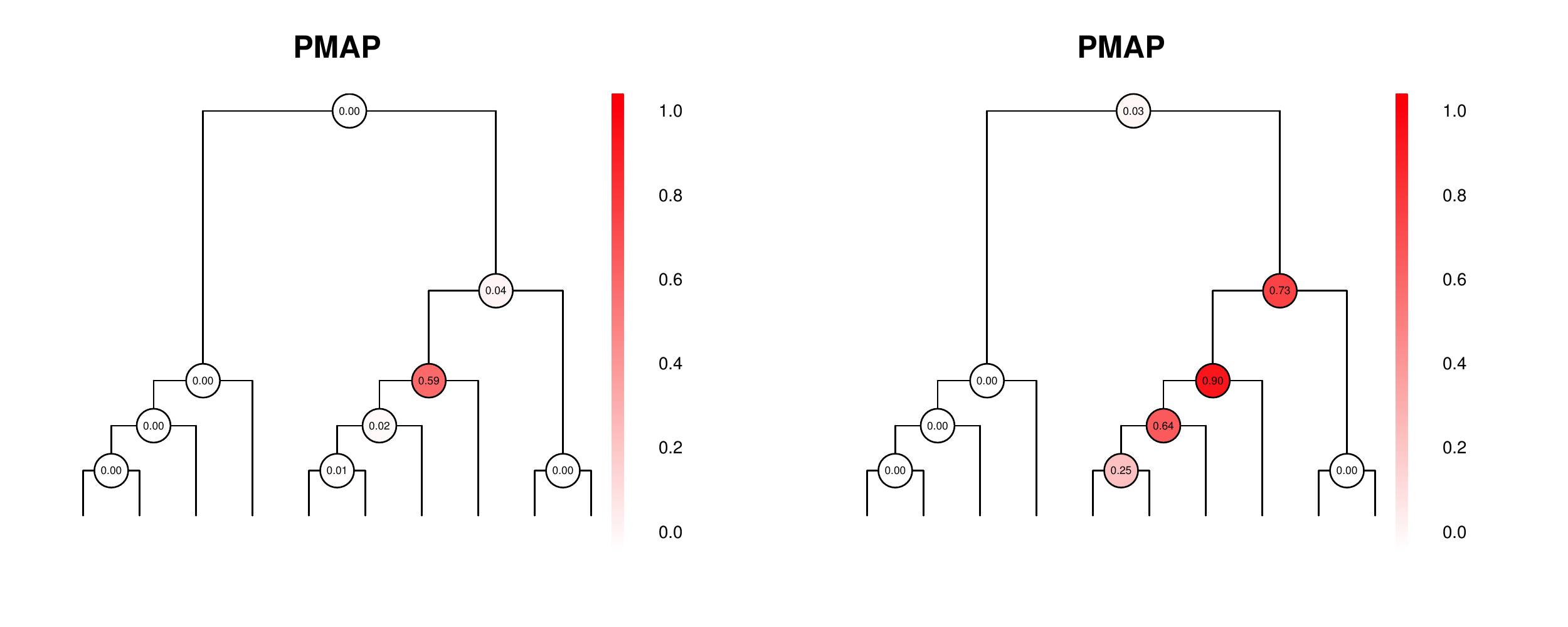}\\
\caption{{\color{black}PMAPs from BCR and BGCR under a specific simulation case in Scenario $\RN{3}$. Only nodes in the zoom-in subtree in \ref{fig:sim3_tree} are shown since the PMAPs at other nodes are essentially zero. Left: PMAPs obtained by BCR; Right: PMAPs obtained by BGCR. The number in each node denotes its PMAP. We also color the nodes by their PMAPs to give a better visualization.}}
\label{fig:sim3_ex}
\end{center}
\end{figure}

In many applications, characterizing {\color{black}and visualizing} the cross-group difference is often of more interests than merely testing its existence. {\color{black} As mentioned in Section \ref{subsec:dec}, this can be achieved by marking the PMAPs along the phylogenetic tree. As an example, consider a specific round of simulation in Scenario \RN{3}. In this simulation, $\text{PJAP}_{\text{BGCR}}=0.910$, $\text{PJAP}_{\text{BCR}}=0.651$, $\hat\gamma = 5.521$. \ref{fig:sim3_ex} plots the PMAPs obtained by the two models along the phylogenetic tree (only shows the relevant part of the tree). It is clear that BGCR benefits from the graphical structure and reveals the designed chaining pattern of the cross-group differences (see \ref{fig:sim3_tree}).}

\subsection{Adjusting for covariates}
\label{subsec:adj}
In practice, OTU compositions can depend on various factors. When the data are gathered from observational studies, 
unadjusted covariates can lead to false positives. Even when the data do arise from randomized experiments, incorporating relevant covariates can 
improve the statistical power for identifying cross-group differences. We illustrate with another simulation scenario the necessity of incorporating confounders in the testing procedure:
\begin{enumerate}
\item [$\RN{4}$.]{\textbf{Cross-group difference exists at a single OTU with an unbalanced confounder.}} We start from the same American Gut dataset as in the previous simulation scenarios. 
Now we randomly select 200 \textit{male} and 200 \textit{female} Caucasians, all middle aged and from the west region (the reason we take a random sample instead of using the full dataset is the large computational burden brought by the competitors). To purposefully make gender {\color{black}an unbalanced} confounder, we select an OTU (OTU `4352657', {\color{black}denoted by $\omega_{c}$}) and increase the counts of this OTU for all male participants by $175\%$. We then randomly select 160 males and 40 females into the first group and put the rest into the second group to get the data under the null. To simulate data under the alternative, we select another OTU (OTU `4481131', {\color{black}denoted by $\omega_{s}$}) and increase the count of this OTU in the second group by $175\%$.   
\end{enumerate}

\begin{figure}[!h]
\begin{center}
\includegraphics[width = 17cm]{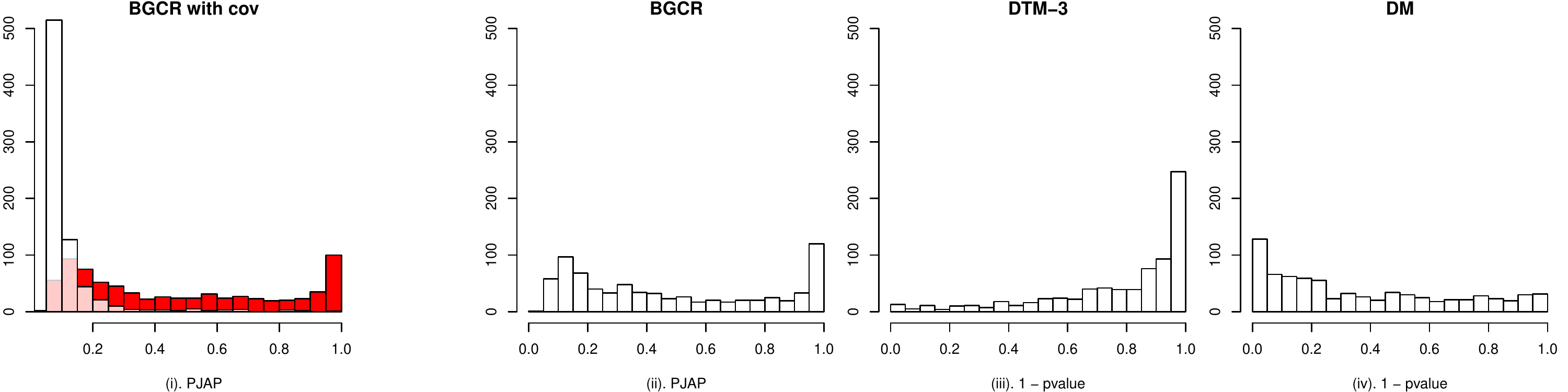}\\
\caption{{\color{black}Histograms of the statistics used for decision making under the null. Plot (i) shows the histogram of PJAPs under the null  (white) and under the alternative (red) with ``gender'' adjusted. The other three plots present the histograms of the decision-making statistics of the three models under comparison under the null without adjusting for ``gender''. } }
\label{fig:sim4_hist}
\end{center}
\end{figure}

We carry out $750$ simulations with $K=50$ {\color{black}and compare BGCR with DTM-3 and DM} (we do not simulate 3000 rounds or consider $K=75$ or $100$ since some competitors are too slow).  {\color{black}\ref{fig:sim4_hist} shows the histograms of the statistics these methods used for decision making under the null---PJAP for BGCR, $p$-values (or equivalently, $1-p$-values ) for DTM-3 and DM. From \ref{fig:sim4_hist} (ii)-(iv), it is clear that all three methods fail in controlling false positives without adjusting for the ``gender'' of the participants. After incorporating ``gender'' in each node-specific regression, BGCR is able to keep the PJAPs under the null at a reasonable level (\ref{fig:sim4_hist} (i))}. 

\begin{figure}[!h]
\begin{center}
\includegraphics[width = 17cm]{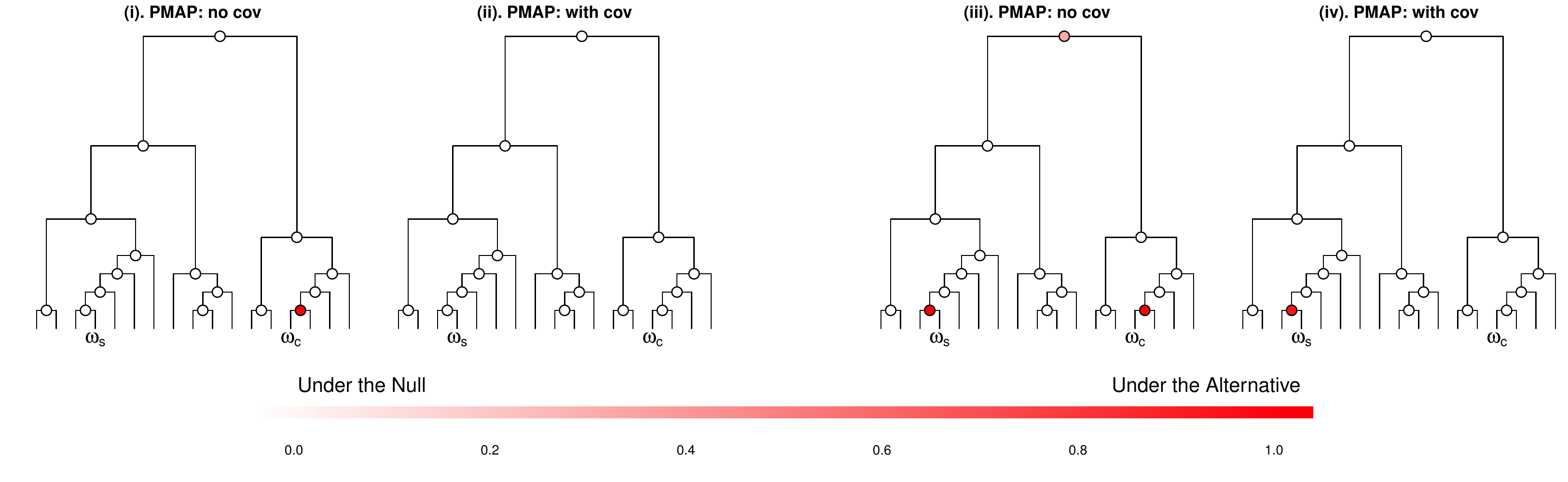}\\
\caption{{\color{black}PMAPs of a specific simulation in Scenario \RN{4}. (i) and (ii): PMAPs under the null; (iii) and (iv): PMAPs under the alternative (only a subtree is shown since PMAPs on other nodes are essentially zero). The nodes are colored by their PMAPs. Under the alternative, true cross-group differences are only present at the parent of $\omega_s$. A false discovery is made at the parent of $\omega_c$ when ``gender'' is not adjusted for.}}
\label{fig:sim4_ex}
\end{center}
\end{figure}

Adjusting for confounders is also important in the characterization of the cross-group differences. {\color{black}For example, consider a specific round of simulation in Scenario $\RN{4}$. \ref{fig:sim4_ex} marks the PMAPs reported by BGCR on the phylogenetic tree under the null and the alternative, with or without ``gender'' included. Under the null, ignoring ``gender'' leads to a false discovery at the parent of $\omega_c$ and also a false rejection of the hypothesis that no cross-group difference exists at threshold $0.95$. Under the alternative, the PJAPs reported by BGCR are closed to $1$ with or without adjusting for ``gender'', suggesting that incorporating ``gender'' in the model is likely to keep the decision about the original hypothesis unchanged. However, failing to adjust for ``gender'' contaminates the pattern of the cross-group differences (\ref{fig:sim4_ex} (iii)-(iv)). }
 
\section{Application to the American Gut project}
\label{sec:app}

\subsection{Cross-group comparison of OTU compositions }
\label{subsec:application}

{\color{black}In this section, we apply BGCR to the same dataset from the American Gut project as in \ref{sec:num_exam}. The American Gut project collects many binary or ordered categorical covariates of the participants that can be used to divide them into groups. In particular, we focus on comparing the OTU compositions of groups of participants categorized by their dietary habits, which can be characterized by their frequencies of consuming different kinds of food. 
 We consider eight types of food: \textit{alcohol, diary, fruit, meat/egg, seafood, sweets, vegetable} and \textit{whole grain}. For each kind of food,} the dataset records each participant's frequency of consumption per week as a categorical variable with five categories: \textit{Never, Occasionally (1-2 times per week), Rarely (less than once per week), Regularly (3-5 times per week) and Daily}. We dichotomize the categories based on whether the food is consumed less than 3 times per week and use this binary variable to divide the participants into two groups.

{\color{black}In our analysis, we only use the top $100$ OTUs with the largest overall counts across samples. This choice is made mainly for quality control. For OTUs containing nil or only very few counts over all samples, their counts are largely influenced by sequencing errors and are highly unreliable. Using these OTUs is likely to contaminate the analysis by bringing in pure noises that are hard to quantify. We note that prescreening the OTUs to drop those who do not have substantial counts in any samples is a common recommended 
practice in the literature (see for example \cite{wu2011linking, tang:2017}). }

\begin{figure}[!h]
\begin{center}
\includegraphics[width = 11cm]{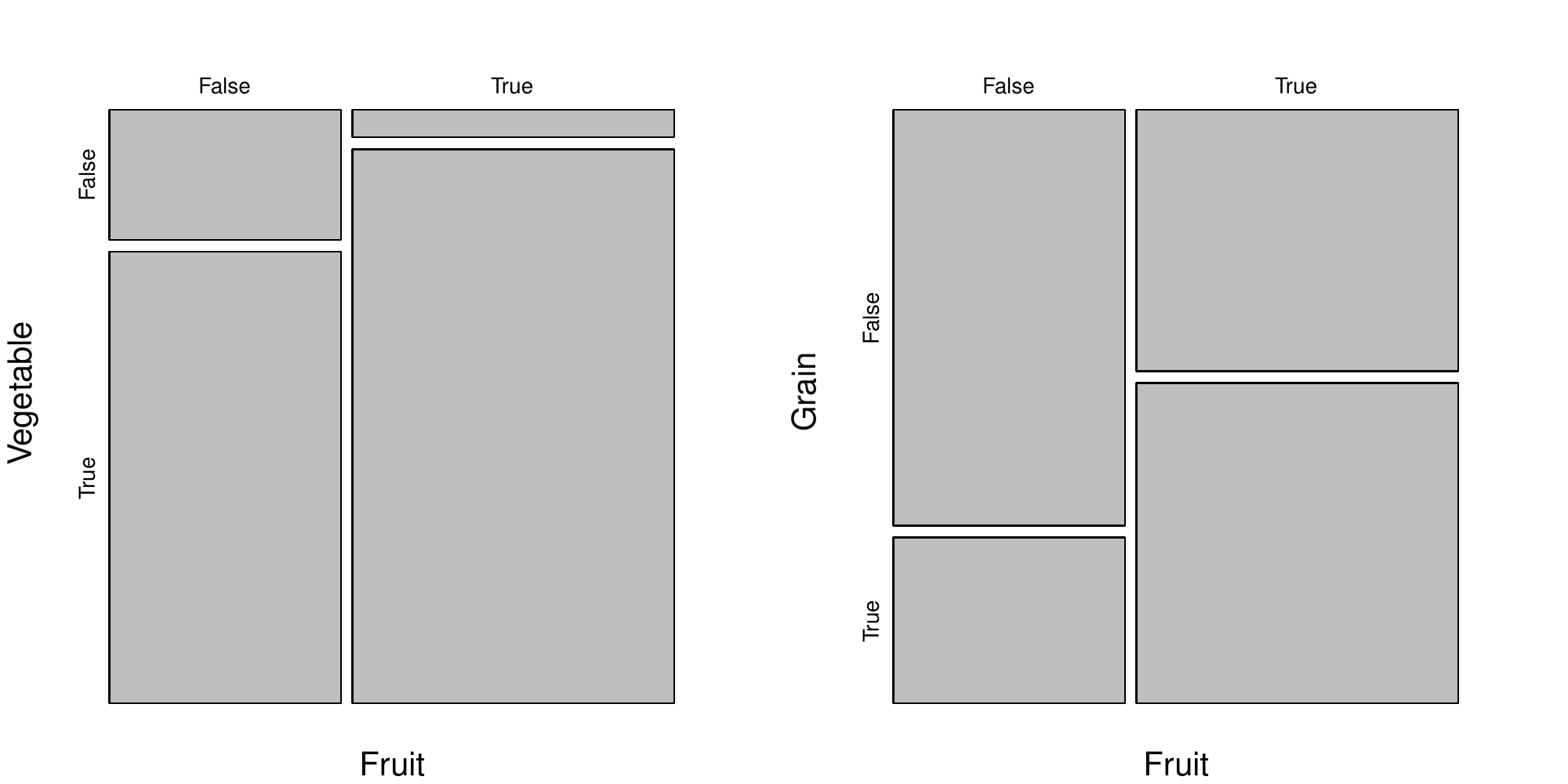}\\
\caption{Unbalancedness of vegetable and grain consumption in the two groups defined by fruit consumption. The area of each block represents the proportion of the counts in the corresponding cell of the $2\times 2$ contingency table.}
\label{fig:app_unbalance}
\end{center}
\end{figure}

Besides diet, there are other possible factors that may influence the OTU composition. Including these factors in the model could help reduce the chance of false discoveries and improve the power of the inference. In our analysis, we consider five such covariates: \textit{the participant's biological sex, has the participant been diagnosed with diabetes, does the participant have IBD (inflammatory bowel disease), has the participant used antibiotic in the past year, and does the participant consume probiotic less than 3 times per week}. All five covariates are binary. {\color{black}In addition to adjusting for these non-dietary covariates when investigating a certain kind of food, dietary habits of other foods are also natural candidates for confounders to be included in the analysis, especially when they are unbalanced across the two groups. For example, \ref{fig:app_unbalance} shows the unbalancedness of vegetable and grain consumption of participants across the two groups defined by fruit consumption. Since the covariate information in the study is collected by asking the participants to take a non-compulsory questionnaire, some covariates are missing for certain participants.} For simplicity, we ignore all participants with missing values in any variable mentioned above and perform a complete-case analysis. Moreover, we restrict our samples to participants with age in range $20\sim 69$ and BMI $18.5\sim 30$ to disregard potential outliers. The resulting sample size in each comparison is shown in \ref{tab:diet}.

\begin{table}[h]
\begin{center}
\begin{tabular}{c | c c | c   c   c  }
\hline
\multirow{2}{*}{Diet}      & \multicolumn{2}{|c|}{Group size} & \multicolumn{3}{c}{$1-$PJAP}  \\ \cline{2-6}
 & False & True & No cov & Non-diet cov & All cov  ($\hat\tau$) \\ \hline
Alcohol & 2099 & 995 &  $0.134$ & $0.345$ &  $0.555$ $(0)$  \\
Diary & 1733 & 1361 & $1.27 \times 10^{-2}$ & $6.97\times 10^{-2}$ & $0.517$  $(0)$  \\
Fruit &  1296 & 1798 &  $6.86\times 10^{-7}$ & $1.29\times 10^{-8}$ & {\color{red}$0.259$} $(3.93)$ \\
Meat/Egg &  798 & 2296 & $0.867$ & $0.857$ & $0.757$ $(0)$  \\
Seafood & 2637 & 457  &  $1.99\times 10^{-2}$ & $9.11\times 10^{-3}$ & {\color{red}$0.166$} $(0)$   \\
Sweets & 2005 & 1089  & $0.200$ & $0.347$ & $0.912$ $(0)$  \\
Vegetable & 375 & 2719 &  $ < 1.00\times 10^{-8}$ & $< 1.00\times 10^{-8}$ &  {\color{red}$< 1.00\times 10^{-8}$} $(5.13)$   \\
Whole grain  & 1735 & 1359 & $ < 1.00\times 10^{-8}$ & $ < 1.00\times 10^{-8}$ &  {\color{red}$3.85\times 10^{-7}$} $(5.63)$    \\
\hline
\end{tabular}
\end{center}
\caption{{\color{black}($1-$PJAP)'s} for testing OTU compositions of different dietary habits. In the last column, red cells indicate rejections of the global null under threshold $L = 0.5$. The estimated $\hat\tau$ when all the covariates are adjusted is shown in the parentheses in the last column.}
\label{tab:diet}
\end{table}

\begin{figure}[!h]
\begin{center}
\includegraphics[width = 14cm]{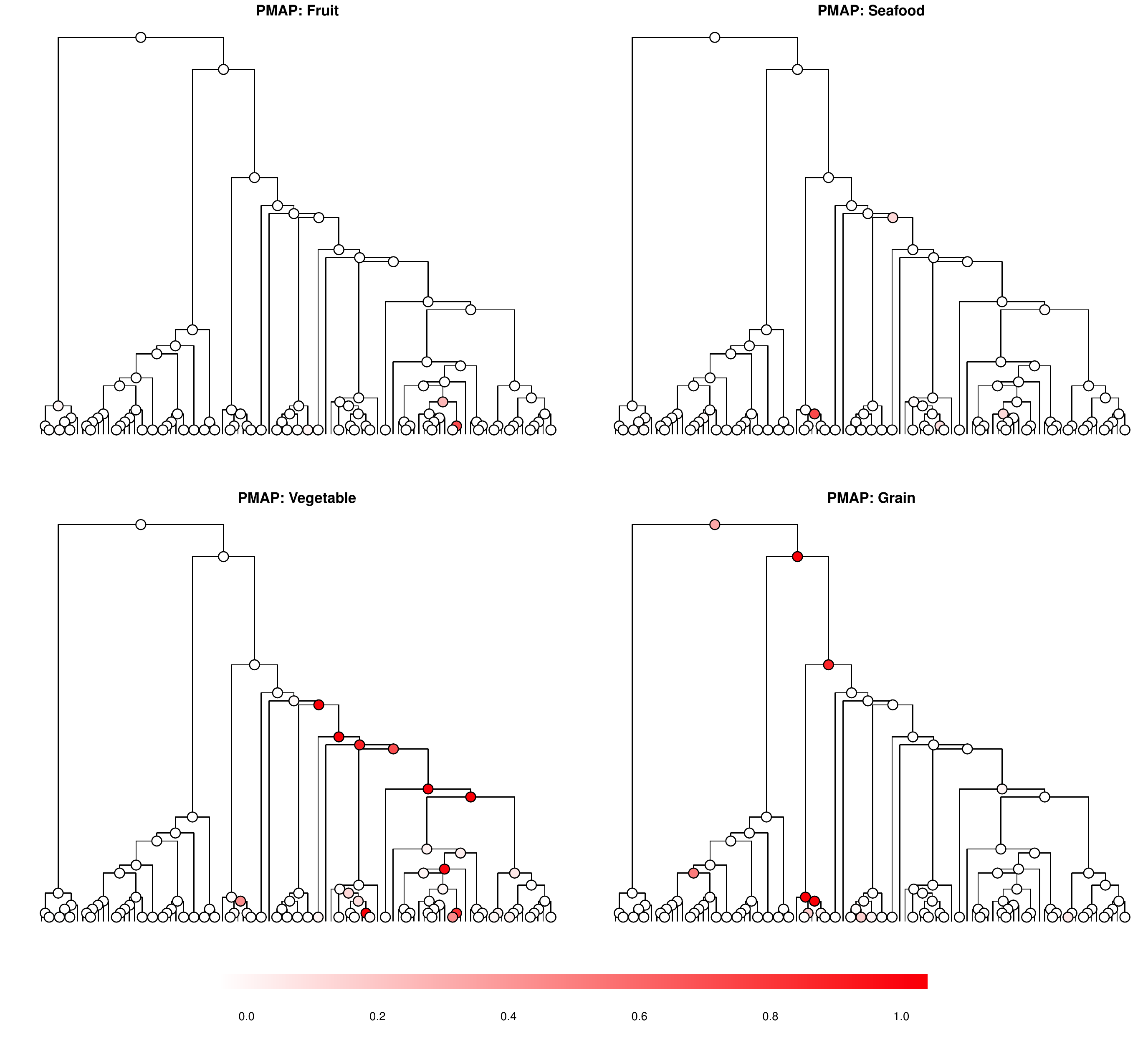}
\caption{{\color{black}PMAPs for the four comparisons that reject the global null. The nodes are colored by PMAPs reported by BGCR with both non-dietary covariates and dietary covariates adjusted.}}
\label{fig:app_pmap}
\end{center}
\end{figure}

{\color{black}For each type of food, we use BGCR to compare the group-level OTU composition across the two resulting groups. In each comparison, we consider BGCR with different sets of covariates adjusted---first no covariate, then the five non-dietary covariates, and finally the non-dietary covariates as well as dietary habits on other foods. The $(1-\text{PJAP})$'s reported by BGCR under each comparison are shown in \ref{tab:diet}.} When all the covariates are adjusted, the test rejects the global null at threshold $L = 0.5$ when the {\color{black}grouping food} is \textit{fruit, seafood, vegetable} or \textit{whole grain}. {\color{black}This is consistent with the findings that dietary habits are closely related to gut microbiome compositions \citep{graf2015contribution,singh2017influence,rothschild2018environment}. Typically, the posterior probability of the null increases when more covariates are incorporated in the model. This reflects the complex nature of the microbial community that it is associated with various factors \citep{rothschild2018environment}. For \textit{diary} and \textit{sweet}, adjusting for the dietary covariates is crucial. When no covariate or only non-dietary covariates are adjusted, the null of no cross-group difference is likely to be rejected. Once the dietary covariates are included, the PJAPs change dramatically and can reverse the testing decision. For the four comparisons that reject the null, we visualize the cross-group differences by marking the PMAPs on the phylogenetic tree (\ref{fig:app_pmap}). Similar plots of PMAPs under BGCR with no covariate or only non-dietary covariates can be found in online supplementary materials (Figure S7 and Figure S8). These plots show that adjusting for relevant covariates greatly changes the characterization of the cross-group differences. Many significant nodes reported by BGCR with some covariates unadjusted turn out to be false positives when all the covariates are included. }

{\color{black}

\subsection{BGCR vs BCR}
\label{subsec:validate}

In this section, we evaluate the performance of BGCR compared with BCR in our data application. In BGCR, the global parameter $\tau$ controls the level of dependencies among the node-specific tests. We note in Section~\ref{subsec:hyper} that when $\tau=0$, BGCR degenerates to BCR. This nested feature together with our empirical Bayes strategy to estimate $\tau$ allow BGCR to decide whether the dependencies are necessary based on the data---when the estimated $\hat\tau = 0$, BGCR simply performs independent tests without introducing the graphical structure.

\ref{tab:diet} shows $\hat\tau$ in each comparison when BGCR with all covariates is used. In all four comparisons that the null is accepted at threshold $
L=0.5$, $\hat\tau=0$. In these cases, no node is significant and the ``signals'' are trivially independent. {\color{black} For \textit{fruit, vegetable} and \textit{whole grain}, $\hat\tau > 0$, suggesting that cross-group differences might cluster into chains. To see whether it is necessary to introduce the graphical structure among the node-specific tests, one can test $H^g_0: \tau = 0$ vs $H^g_1: \tau\not=0$. For example, when the prior $\tau\sim\text{Uniform}(0, 6)$ is used, the Bayes factor comparing $H^g_1$ and $H^g_0$ for \textit{fruit, vegetable, seafood} and \textit{whole grain} are $1.022$, $0.830$, $138.699$ and $6.922$, respectively. For \textit{vegetable} and \textit{whole grain}, the cross-group differences cluster into chains (\ref{fig:app_pmap}) and thus it is necessary to introduce the graphical structure. We can also look at the posterior on $\tau$ in the four comparisons that reject the null as shown in \ref{fig:app_post}. For \textit{vegetable} and \textit{whole grain}, the signals demonstrate a strong chaining pattern and the posteriors on $\tau$ concentrate to some positive values. For \textit{fruit} and \textit{seafood}, the cross-group differences are local to certain nodes. Therefore, not much information about the dependencies of the node-specific tests is provided by the data and the posteriors on $\tau$ are very similar to the prior. }

\begin{figure}[!h]
\begin{center}
\includegraphics[width = 17cm]{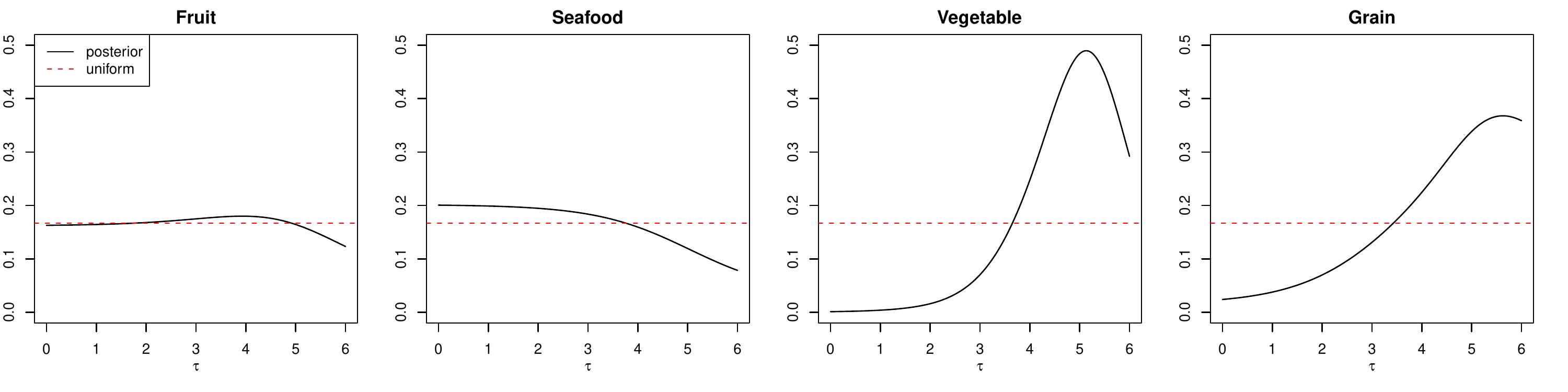}
\caption{Posteriors on $\tau$ with the uniform prior. Shown for the four comparisons that reject the null at threshold $L=0.5$.}
\label{fig:app_post}
\end{center}
\end{figure}
}

\vspace{-7mm}
\section{Concluding remarks}
\label{sec:discussion}

We have introduced a Bayesian framework for testing the existence of cross-group differences in the OTU composition. The {\color{black}original} testing problem is transformed into a series of dependent {\color{black}node-specific} tests linked together by the phylogenetic tree. A key feature of the Bayesian formulation is that the dispersion parameters are integrated out, providing more robust inference than the methods that use point estimates of the within-group variation. We use numerical integration to evaluate the Bayes factors, circumventing Monte Carlo simulations to give fast inference. By introducing dependencies among the local tests, information sharing is allowed among neighboring nodes. This further improves the power of the test when the cross-group differences cluster into chains along the tree, which is often observed in practice. We derive an exact message passing algorithm that is scalable to the size of the tree to carry out inference under this information borrowing.

{\color{black}Because microbiome data are often collected by observational studies, it is important to adjust for possible confounders in testing. This is crucial in reducing the chance of false discoveries and the residual overdispersion.} Our model achieves this goal by incorporating the covariates via a regression model, which adds little computational burdens to the Bayesian testing framework. {\color{black}A relevant problem is to select covariates in the testing scenario. Generally, this must be done with caution since incorporating standard variable selection procedures can substantially affect or even invalidate the meaning of the testing result on the two-group difference due to possible confounding. Readers may refer to Online Supplementary Materials~C for more details on this along with some numerical examples.}

Besides testing, our model gives a full probabilistic characterization of the cross-group differences that can be naturally visualized. Note that although we focus on two group comparisons in this paper, the proposed method can be generalized to multiple group settings with little extra effort.

 \vspace{-7mm}
 \section*{Software}
 The DTM methods and the DM test are implemented with the \texttt{R} packages \texttt{PhyloScan} and \texttt{HMP}. \texttt{R} code for BGCR is freely available at \url{https://github.com/MaStatLab/BGCR}. {\color{black}Source code for the numerical examples are available at \url{https://github.com/MaStatLab/BGCR_analysis}}. 
\vspace{-7mm}
 \section*{Acknowledgment}
 LM's research is partly supported by NSF grants DMS-1612889, DMS-1749789, and a Google Faculty Research Award. Part of this work was completed when JM was supported by a Duke Forge Fellowship.
 
\bibliography{DM}

\begin{thebibliography}{}

\bibitem[\protect\citeauthoryear{Barbieri, Berger, et~al.}{Barbieri
  et~al.}{2004}]{barbieri2004optimal}
Barbieri, M.~M., J.~O. Berger, et~al. (2004).
\newblock Optimal predictive model selection.
\newblock {\em The annals of statistics\/}~{\em 32\/}(3), 870--897.

\bibitem[\protect\citeauthoryear{Berger, Pericchi, Ghosh, Samanta, De~Santis,
  Berger, and Pericchi}{Berger et~al.}{2001}]{berger2001objective}
Berger, J.~O., L.~R. Pericchi, J.~Ghosh, T.~Samanta, F.~De~Santis, J.~Berger,
  and L.~Pericchi (2001).
\newblock Objective bayesian methods for model selection: introduction and
  comparison.
\newblock {\em Lecture Notes-Monograph Series\/}, 135--207.

\bibitem[\protect\citeauthoryear{Callahan, McMurdie, and Holmes}{Callahan
  et~al.}{2017}]{callahan2017exact}
Callahan, B.~J., P.~J. McMurdie, and S.~P. Holmes (2017).
\newblock Exact sequence variants should replace operational taxonomic units in
  marker-gene data analysis.
\newblock {\em The ISME journal\/}~{\em 11\/}(12), 2639.

\bibitem[\protect\citeauthoryear{Caporaso, Kuczynski, Stombaugh, Bittinger,
  Bushman, Costello, Fierer, Pe{\~n}a, Goodrich, Gordon, et~al.}{Caporaso
  et~al.}{2010}]{caporaso2010qiime}
Caporaso, J.~G., J.~Kuczynski, J.~Stombaugh, K.~Bittinger, F.~D. Bushman, E.~K.
  Costello, N.~Fierer, A.~G. Pe{\~n}a, J.~K. Goodrich, J.~I. Gordon, et~al.
  (2010).
\newblock Qiime allows analysis of high-throughput community sequencing data.
\newblock {\em Nature methods\/}~{\em 7\/}(5), 335--336.

\bibitem[\protect\citeauthoryear{Chen and Friedman}{Chen and
  Friedman}{2017}]{chen2017new}
Chen, H. and J.~H. Friedman (2017).
\newblock A new graph-based two-sample test for multivariate and object data.
\newblock {\em Journal of the American statistical association\/}~{\em
  112\/}(517), 397--409.

\bibitem[\protect\citeauthoryear{Chen and Li}{Chen and
  Li}{2013}]{chen2013variable}
Chen, J. and H.~Li (2013).
\newblock Variable selection for sparse dirichlet-multinomial regression with
  an application to microbiome data analysis.
\newblock {\em The annals of applied statistics\/}~{\em 7\/}(1).

\bibitem[\protect\citeauthoryear{Crouse, Nowak, and Baraniuk}{Crouse
  et~al.}{1998}]{crouse1998markov}
Crouse, M.~S., R.~D. Nowak, and R.~G. Baraniuk (1998).
\newblock Wavelet-based statistical signal processing using hidden markov
  models.
\newblock {\em IEEE Transactions on signal processing\/}~{\em 46\/}(4),
  886--902.

\bibitem[\protect\citeauthoryear{David, Maurice, Carmody, Gootenberg, Button,
  Wolfe, Ling, Devlin, Varma, Fischbach, et~al.}{David
  et~al.}{2014}]{david2014diet}
David, L.~A., C.~F. Maurice, R.~N. Carmody, D.~B. Gootenberg, J.~E. Button,
  B.~E. Wolfe, A.~V. Ling, A.~S. Devlin, Y.~Varma, M.~A. Fischbach, et~al.
  (2014).
\newblock Diet rapidly and reproducibly alters the human gut microbiome.
\newblock {\em Nature\/}~{\em 505\/}(7484), 559--563.

\bibitem[\protect\citeauthoryear{Dennis}{Dennis}{1991}]{dennis1991hyper}
Dennis, S.~Y. (1991).
\newblock On the hyper-dirichlet type 1 and hyper-liouville distributions.
\newblock {\em Communications in Statistics-Theory and Methods\/}~{\em
  20\/}(12), 4069--4081.

\bibitem[\protect\citeauthoryear{Dennis}{Dennis}{1996}]{dennis1996bayesian}
Dennis, S.~Y. (1996).
\newblock A bayesian analysis of tree-structured statistical decision problems.
\newblock {\em Journal of statistical planning and inference\/}~{\em 53\/}(3),
  323--344.

\bibitem[\protect\citeauthoryear{George and McCulloch}{George and
  McCulloch}{1997}]{george1997approaches}
George, E.~I. and R.~E. McCulloch (1997).
\newblock Approaches for bayesian variable selection.
\newblock {\em Statistica sinica\/}, 339--373.

\bibitem[\protect\citeauthoryear{Graf, Di~Cagno, F{\aa}k, Flint, Nyman,
  Saarela, and Watzl}{Graf et~al.}{2015}]{graf2015contribution}
Graf, D., R.~Di~Cagno, F.~F{\aa}k, H.~J. Flint, M.~Nyman, M.~Saarela, and
  B.~Watzl (2015).
\newblock Contribution of diet to the composition of the human gut microbiota.
\newblock {\em Microbial ecology in health and disease\/}~{\em 26\/}(1), 26164.

\bibitem[\protect\citeauthoryear{Grantham, Reich, Borer, and Gross}{Grantham
  et~al.}{2017}]{grantham2017mimixf}
Grantham, N.~S., B.~J. Reich, E.~T. Borer, and K.~Gross (2017).
\newblock Mimix: a bayesian mixed-effects model for microbiome data from
  designed experiments.
\newblock {\em arXiv preprint arXiv:1703.07747\/}.

\bibitem[\protect\citeauthoryear{Held, Bov{\'e}, Gravestock, et~al.}{Held
  et~al.}{2015}]{held2015approximate}
Held, L., D.~S. Bov{\'e}, I.~Gravestock, et~al. (2015).
\newblock Approximate bayesian model selection with the deviance statistic.
\newblock {\em Statistical Science\/}~{\em 30\/}(2), 242--257.

\bibitem[\protect\citeauthoryear{Hildebrandt, Hoffmann, Sherrill-Mix,
  Keilbaugh, Hamady, Chen, Knight, Ahima, Bushman, and Wu}{Hildebrandt
  et~al.}{2009}]{hildebrandt2009high}
Hildebrandt, M.~A., C.~Hoffmann, S.~A. Sherrill-Mix, S.~A. Keilbaugh,
  M.~Hamady, Y.-Y. Chen, R.~Knight, R.~S. Ahima, F.~Bushman, and G.~D. Wu
  (2009).
\newblock High-fat diet determines the composition of the murine gut microbiome
  independently of obesity.
\newblock {\em Gastroenterology\/}~{\em 137\/}(5), 1716--1724.

\bibitem[\protect\citeauthoryear{Holmes, Caron, Griffin, Stephens,
  et~al.}{Holmes et~al.}{2015}]{holmes2015two}
Holmes, C.~C., F.~Caron, J.~E. Griffin, D.~A. Stephens, et~al. (2015).
\newblock Two-sample bayesian nonparametric hypothesis testing.
\newblock {\em Bayesian Analysis\/}~{\em 10\/}(2), 297--320.

\bibitem[\protect\citeauthoryear{Koller and Friedman}{Koller and
  Friedman}{2009}]{koller2009probabilistic}
Koller, D. and N.~Friedman (2009).
\newblock {\em Probabilistic graphical models: principles and techniques}.
\newblock MIT press.

\bibitem[\protect\citeauthoryear{La~Rosa, Brooks, Deych, Boone, Edwards, Wang,
  Sodergren, Weinstock, and Shannon}{La~Rosa et~al.}{2012}]{la:2012}
La~Rosa, P.~S., J.~P. Brooks, E.~Deych, E.~L. Boone, D.~J. Edwards, Q.~Wang,
  E.~Sodergren, G.~Weinstock, and W.~D. Shannon (2012).
\newblock Hypothesis testing and power calculations for taxonomic-based human
  microbiome data.
\newblock {\em PloS one\/}~{\em 7\/}(12), e52078.

\bibitem[\protect\citeauthoryear{Lauritzen and Spiegelhalter}{Lauritzen and
  Spiegelhalter}{1988}]{lauritzen1988local}
Lauritzen, S.~L. and D.~J. Spiegelhalter (1988).
\newblock Local computations with probabilities on graphical structures and
  their application to expert systems.
\newblock {\em Journal of the Royal Statistical Society. Series B
  (Methodological)\/}, 157--224.

\bibitem[\protect\citeauthoryear{Li}{Li}{2015}]{li2015microbiome}
Li, H. (2015).
\newblock Microbiome, metagenomics, and high-dimensional compositional data
  analysis.
\newblock {\em Annual Review of Statistics and Its Application\/}~{\em 2},
  73--94.

\bibitem[\protect\citeauthoryear{Li and Clyde}{Li and
  Clyde}{2015}]{li2015mixtures}
Li, Y. and M.~A. Clyde (2015).
\newblock Mixtures of g-priors in generalized linear models.
\newblock {\em arXiv preprint arXiv:1503.06913\/}.

\bibitem[\protect\citeauthoryear{Liang, Paulo, Molina, Clyde, and Berger}{Liang
  et~al.}{2008}]{liang2008mixtures}
Liang, F., R.~Paulo, G.~Molina, M.~A. Clyde, and J.~O. Berger (2008).
\newblock Mixtures of g priors for bayesian variable selection.
\newblock {\em Journal of the American Statistical Association\/}~{\em
  103\/}(481), 410--423.

\bibitem[\protect\citeauthoryear{Ma and Soriano}{Ma and
  Soriano}{2018}]{ma:2016}
Ma, L. and J.~Soriano (2018).
\newblock Analysis of distributional variation through graphical multi-scale
  beta-binomial models.
\newblock {\em Journal of Computational and Graphical Statistics\/}, 1--13.

\bibitem[\protect\citeauthoryear{McDonald, Hornig, Lozupone, Debelius, Gilbert,
  and Knight}{McDonald et~al.}{2015}]{mcdonald2015towards}
McDonald, D., M.~Hornig, C.~Lozupone, J.~Debelius, J.~A. Gilbert, and R.~Knight
  (2015).
\newblock Towards large-cohort comparative studies to define the factors
  influencing the gut microbial community structure of asd patients.
\newblock {\em Microbial ecology in health and disease\/}~{\em 26}.

\bibitem[\protect\citeauthoryear{M{\"u}ller, Parmigiani, and Rice}{M{\"u}ller
  et~al.}{2006}]{muller2006fdr}
M{\"u}ller, P., G.~Parmigiani, and K.~Rice (2006).
\newblock Fdr and bayesian multiple comparisons rules.

\bibitem[\protect\citeauthoryear{M{\"u}ller, Parmigiani, Robert, and
  Rousseau}{M{\"u}ller et~al.}{2004}]{muller2004optimal}
M{\"u}ller, P., G.~Parmigiani, C.~Robert, and J.~Rousseau (2004).
\newblock Optimal sample size for multiple testing: the case of gene expression
  microarrays.
\newblock {\em Journal of the American Statistical Association\/}~{\em
  99\/}(468), 990--1001.

\bibitem[\protect\citeauthoryear{Qin, Li, Cai, Li, Zhu, Zhang, Liang, Zhang,
  Guan, Shen, et~al.}{Qin et~al.}{2012}]{qin2012metagenome}
Qin, J., Y.~Li, Z.~Cai, S.~Li, J.~Zhu, F.~Zhang, S.~Liang, W.~Zhang, Y.~Guan,
  D.~Shen, et~al. (2012).
\newblock A metagenome-wide association study of gut microbiota in type 2
  diabetes.
\newblock {\em Nature\/}~{\em 490\/}(7418), 55--60.

\bibitem[\protect\citeauthoryear{Ren, Bacallado, Favaro, Vatanen, Huttenhower,
  and Trippa}{Ren et~al.}{2017}]{ren2017bayesian}
Ren, B., S.~Bacallado, S.~Favaro, T.~Vatanen, C.~Huttenhower, and L.~Trippa
  (2017).
\newblock Bayesian nonparametric mixed effects models in microbiome data
  analysis.
\newblock {\em arXiv preprint arXiv:1711.01241\/}.

\bibitem[\protect\citeauthoryear{Rothschild, Weissbrod, Barkan, Kurilshikov,
  Korem, Zeevi, Costea, Godneva, Kalka, Bar, et~al.}{Rothschild
  et~al.}{2018}]{rothschild2018environment}
Rothschild, D., O.~Weissbrod, E.~Barkan, A.~Kurilshikov, T.~Korem, D.~Zeevi,
  P.~I. Costea, A.~Godneva, I.~N. Kalka, N.~Bar, et~al. (2018).
\newblock Environment dominates over host genetics in shaping human gut
  microbiota.
\newblock {\em Nature\/}~{\em 555\/}(7695), 210.

\bibitem[\protect\citeauthoryear{Singh, Chang, Yan, Lee, Ucmak, Wong, Abrouk,
  Farahnik, Nakamura, Zhu, et~al.}{Singh et~al.}{2017}]{singh2017influence}
Singh, R.~K., H.-W. Chang, D.~Yan, K.~M. Lee, D.~Ucmak, K.~Wong, M.~Abrouk,
  B.~Farahnik, M.~Nakamura, T.~H. Zhu, et~al. (2017).
\newblock Influence of diet on the gut microbiome and implications for human
  health.
\newblock {\em Journal of translational medicine\/}~{\em 15\/}(1), 73.

\bibitem[\protect\citeauthoryear{Sogin, Morrison, Huber, Welch, Huse, Neal,
  Arrieta, and Herndl}{Sogin et~al.}{2006}]{sogin2006microbial}
Sogin, M.~L., H.~G. Morrison, J.~A. Huber, D.~M. Welch, S.~M. Huse, P.~R. Neal,
  J.~M. Arrieta, and G.~J. Herndl (2006).
\newblock Microbial diversity in the deep sea and the underexplored ``rare
  biosphere''.
\newblock {\em Proceedings of the National Academy of Sciences\/}~{\em
  103\/}(32), 12115--12120.

\bibitem[\protect\citeauthoryear{Soriano and Ma}{Soriano and
  Ma}{2017}]{soriano2017probabilistic}
Soriano, J. and L.~Ma (2017).
\newblock Probabilistic multi-resolution scanning for two-sample differences.
\newblock {\em Journal of the Royal Statistical Society: Series B (Statistical
  Methodology)\/}~{\em 79\/}(2), 547--572.

\bibitem[\protect\citeauthoryear{Tang, Ma, Nicolae, et~al.}{Tang
  et~al.}{2018}]{tang:2017}
Tang, Y., L.~Ma, D.~L. Nicolae, et~al. (2018).
\newblock A phylogenetic scan test on a dirichlet-tree multinomial model for
  microbiome data.
\newblock {\em The Annals of Applied Statistics\/}~{\em 12\/}(1), 1--26.

\bibitem[\protect\citeauthoryear{Tang and Nicolae}{Tang and
  Nicolae}{2017}]{tang2017mixed}
Tang, Y. and D.~L. Nicolae (2017).
\newblock Mixed effect dirichlet-tree multinomial for longitudinal microbiome
  data and weight prediction.
\newblock {\em arXiv preprint arXiv:1706.06380\/}.

\bibitem[\protect\citeauthoryear{Turnbaugh, Ley, Mahowald, Magrini, Mardis, and
  Gordon}{Turnbaugh et~al.}{2006}]{turnbaugh2006obesity}
Turnbaugh, P.~J., R.~E. Ley, M.~A. Mahowald, V.~Magrini, E.~R. Mardis, and
  J.~I. Gordon (2006).
\newblock An obesity-associated gut microbiome with increased capacity for
  energy harvest.
\newblock {\em nature\/}~{\em 444\/}(7122), 1027--131.

\bibitem[\protect\citeauthoryear{Wadsworth, Argiento, Guindani, Galloway-Pena,
  Shelburne, and Vannucci}{Wadsworth et~al.}{2017}]{wadsworth2017integrative}
Wadsworth, W.~D., R.~Argiento, M.~Guindani, J.~Galloway-Pena, S.~A. Shelburne,
  and M.~Vannucci (2017).
\newblock An integrative bayesian dirichlet-multinomial regression model for
  the analysis of taxonomic abundances in microbiome data.
\newblock {\em BMC bioinformatics\/}~{\em 18\/}(1), 94.

\bibitem[\protect\citeauthoryear{Wang and Zhao}{Wang and
  Zhao}{2017}]{wang2017dirichlet}
Wang, T. and H.~Zhao (2017).
\newblock A dirichlet-tree multinomial regression model for associating dietary
  nutrients with gut microorganisms.
\newblock {\em Biometrics\/}.

\bibitem[\protect\citeauthoryear{Wellman and Henrion}{Wellman and
  Henrion}{1993}]{wellman1993explaining}
Wellman, M.~P. and M.~Henrion (1993).
\newblock Explaining `explaining away'.
\newblock {\em IEEE Transactions on Pattern Analysis and Machine
  Intelligence\/}~{\em 15\/}(3), 287--292.

\bibitem[\protect\citeauthoryear{Wu, Chen, Hoffmann, Bittinger, Chen,
  Keilbaugh, Bewtra, Knights, Walters, Knight, et~al.}{Wu
  et~al.}{2011}]{wu2011linking}
Wu, G.~D., J.~Chen, C.~Hoffmann, K.~Bittinger, Y.-Y. Chen, S.~A. Keilbaugh,
  M.~Bewtra, D.~Knights, W.~A. Walters, R.~Knight, et~al. (2011).
\newblock Linking long-term dietary patterns with gut microbial enterotypes.
\newblock {\em Science\/}~{\em 334\/}(6052), 105--108.

\bibitem[\protect\citeauthoryear{Xia, Chen, Fung, and Li}{Xia
  et~al.}{2013}]{xia2013logistic}
Xia, F., J.~Chen, W.~K. Fung, and H.~Li (2013).
\newblock A logistic normal multinomial regression model for microbiome
  compositional data analysis.
\newblock {\em Biometrics\/}~{\em 69\/}(4), 1053--1063.

\end{thebibliography}


\clearpage
 \appendix
 
\pagenumbering{arabic}
\renewcommand*{\thepage}{S\arabic{page}}
 
 \beginsupplement
 
\section*{Supplementary Materials}

\renewcommand{\thesubsection}{\Alph{subsection}}

\renewcommand{\thefigure}{S\arabic{figure}}

\subsection{Computational strategies }
\label{subsec:comp}

The crux of doing inference with BCR is to calculate the marginal likelihoods $M_s(A)$. We use the computational strategies in \cite{ma:2016} to compute the integrals. Specifically, for fixed $\nu$, the inner integrals on the regression parameters are evaluated based on Laplace approximation; then the outer integral on $\nu$ is calculated with finite Riemann approximation. 

With a bit abuse of the notations, in this section, we use $\bbet(A)$ to denote the `active' regression coefficient in (\ref{eq:hierarchical}). That is, under the null, $\bbet(A)$ is just the $\bbet(A)$ in  (\ref{eq:hierarchical}); under the alternative, $\bbet(A)$ denotes $(\bbet(A)^\top,\gamma(A))^\top$. $\bx_{ij}$ is redefined to be the `active' covariates in the same sense. With these notations, we have $g(\theta_{\bx_{ij}}(A)) =\bx_{ij}^\top\bbet(A)$ and $ \theta_{ij}(A)\mid \bx_{ij}, \nu(A) \sim \mathrm{Beta}(\theta_{\bx_{ij}}(A)\nu(A), (1-\theta_{\bx_{ij}}(A))\nu(A))$ for each local beta-binomial model. 

The computational strategies are the same under both hypotheses. Let $\pi_A(\bbet)$ be the prior density of $\bbet(A)$ under either hypothesis. For fixed $\nu$ in the support of $G_A(\nu)$, by the Laplace approximation, the inner integral is 
\begin{equation*}
\begin{aligned}
L_{\nu}(A)&=\int\prod\limits^{2}_{i=1}\prod\limits^{n_i}_{j=1} \mL_{BB}(g^{-1}(\bx_{ij}^\top\bbet),\nu \mid y_{ij}(A_l), y_{ij}(A_r))\pi_A(\bbet)d\bbet\\
&=\int\exp\left\{ \sum\limits_{i,j} \log  \mL_{BB}(g^{-1}(\bx_{ij}^\top\bbet),\nu \mid y_{ij}(A_l), y_{ij}(A_r))+ \log\pi_A(\bbet)\right\}d\bbet \\
&= \int\exp \left\{ h_\nu(\bbet) \right\} d\bbet \\
&\approx \exp\left\{ h_\nu(\hat\bbet_{\nu}) \right\} \cdot (2\pi)^{d/2}\cdot | -H_{\nu}(\hat\bbet_{\nu})|^{-1/2}\\
&=\hat L_{\nu}(A)
\end{aligned}
\end{equation*}
where $h_\nu(\bbet)= \sum_{i,j} \log  \mL_{BB}(g^{-1}(\bx_{ij}^\top\bbet),\nu \mid y_{ij}(A_l), y_{ij}(A_r)) + \log\pi_A(\bbet)$, $\hat\bbet_{\nu}$ is the maximizer of $h_\nu(\bbet)$, $H_{\nu}(\hat\bbet_{\nu})$ is the Hessian matrix of $h_\nu(\bbet)$ at $\bbet=\hat\bbet_{\nu}$. $d$ is the degrees of freedom of $\bbet$, which is $(p+2)$ under the alternative and $(p+1)$ under the null. We describe a Newton-Raphson algorithm to solve for $\hat\bbet_{\nu}$ below. The log-likelihood function is strictly log-concave and the Newton-Raphson method generally converges after only a few iterations. Finally, to get $M_s(A)$, we compute the outer integral on $\nu$, $\int L_{\nu}(A)dG_A(\nu)$, with finite Riemann approximations. Specifically, after calculating $\hat L_{\nu}(A)$ at a grid of $\nu$'s: $\nu_1,\nu_2,\ldots, \nu_M$, we have
\begin{equation*}
\begin{aligned}
\int L_{\nu}(A)dG_A(\nu)\approx \sum\limits^{M}_{m=2} \hat L_{\nu_m}(A)(G_A(\nu_m)-G_A(\nu_{m-1})).
\end{aligned}
\end{equation*}

\textit{Newton-Rhaphson for $\hat\bbet_{\nu}$}. In this subsection, we shall fix our attention on a specific node $A$ and suppress the `(A)' in the notations. Moreover, we let $y_{ij}(A_l)=t_{ij}$ for simplicity and express the local Beta-Binomial regression model on $A$ as
\begin{equation*}
\begin{aligned}
t_{ij} \sim \mathrm{Binomial}(y_{ij},\theta_{ij}),\quad \theta_{ij}\sim\mathrm{Beta}(\theta_{\bx_{ij}}\nu, (1-\theta_{\bx_{ij}})\nu),\quad \text{and}\quad g(\theta_{\bx})=\bx^\top\bbet.
\end{aligned}
\end{equation*}
The contribution to the log marginal likelihood from the $j$-th observation in group $i$ is
\begin{equation*}
\begin{aligned}
l_{ij}&=\log \mL_{BB}(g^{-1}(\bx_{ij}^\top\bbet),\nu \mid t_{ij}, y_{ij}-t_{ij})\\
&= \log \Gamma(\theta_{\bx_{ij}}\nu+t_{ij}) + \log\Gamma((1-\theta_{\bx_{ij}})\nu+y_{ij}-t_{ij}) - \log\Gamma(\nu+y_{ij}) \\
&\hspace{5mm} -\log\Gamma(\theta_{\bx_{ij}}\nu)-\log\Gamma((1-\theta_{\bx_{ij}})\nu) + \log\Gamma(\nu).
\end{aligned}
\end{equation*}
Taking the first derivative w.r.t. $\bbet$,
\begin{equation*}
\begin{aligned}
\frac{\partial l_{ij}}{\partial\bbet}=\frac{\partial l_{ij}}{\partial\theta_{\bx_{ij}}}\cdot\frac{\partial\theta_{\bx_{ij}}}{\partial\eta_{ij}}\cdot\frac{\partial\eta_{ij}}{\partial\bbet}
\end{aligned}
\end{equation*}
where $\eta_{ij}=\bx_{ij}^\top\bbet$. Now with $\phi$ denoting the digamma function, 
\begin{equation*}
\begin{aligned}
\frac{\partial l_{ij}}{\partial\theta_{\bx_{ij}}}=\nu[ \phi(\theta_{\bx_{ij}}\nu+t_{ij}) - \phi((1-\theta_{\bx_{ij}})\nu+y_{ij}-t_{ij} ) -\phi(\theta_{\bx_{ij}}\nu) + \phi((1-\theta_{\bx_{ij}})\nu) ].
\end{aligned}
\end{equation*}
With the logit link, $\theta_{\bx_{ij}}=g^{-1}(\eta_{ij})=1/(1+e^{-\eta_{ij}})$, and 
\begin{equation*}
\begin{aligned}
\frac{\partial\theta_{\bx_{ij}}}{\partial\eta_{ij}} = (g^{-1})^\prime(\eta_{ij}) = \theta_{\bx_{ij}} (1-\theta_{\bx_{ij}} ).
\end{aligned}
\end{equation*}
Thus
\begin{equation*}
\begin{aligned}
\frac{\partial l_{ij}}{\partial\bbet}=\nu\theta_{\bx_{ij}} (1-\theta_{\bx_{ij}} )[ \phi(\theta_{\bx_{ij}}\nu+t_{ij}) - \phi((1-\theta_{\bx_{ij}})\nu+y_{ij}-t_{ij} ) -\phi(\theta_{\bx_{ij}}\nu) + \phi((1-\theta_{\bx_{ij}})\nu) ]\bx_{ij}.
\end{aligned}
\end{equation*}
The second derivative of $l_{ij}$ w.r.t. $\bbet$ is
{\small\begin{equation*}
\begin{aligned}
\frac{\partial^2 l_{ij}}{\partial\bbet\partial\bbet^\top}= \frac{\partial^2 l_{ij}}{\partial\theta_{\bx_{ij}}^2}\cdot \left(\frac{\partial\theta_{\bx_{ij}}}{\partial\eta_{ij}} \right)^2\cdot \left(\frac{\partial\eta_{ij}}{\partial\bbet} \right)\left( \frac{\partial\eta_{ij}}{\partial\bbet^\top}\right) + \frac{\partial l_{ij}}{\partial\theta_{\bx_{ij}}}\cdot \frac{\partial^2\theta_{\bx_{ij}}}{\partial\eta_{ij}^2}\cdot \left(\frac{\partial\eta_{ij}}{\partial\bbet} \right)\left( \frac{\partial\eta_{ij}}{\partial\bbet^\top}\right) + \frac{\partial l_{ij}}{\partial\theta_{\bx_{ij}}}\cdot\frac{\partial\theta_{\bx_{ij}}}{\partial\eta_{ij}}\frac{\partial^2 \eta_{ij}}{\partial\bbet\partial\bbet^\top}.
\end{aligned}
\end{equation*}  }
The third term on the right-hand side is equal to zero. With $\psi$ being the trigamma function, 
\begin{equation*}
\begin{aligned}
\frac{\partial^2 l_{ij}}{\partial\theta_{\bx_{ij}}^2}=\nu^2[ \psi(\theta_{\bx_{ij}}\nu+t_{ij}) + \psi((1-\theta_{\bx_{ij}})\nu+y_{ij}-t_{ij} ) -\psi(\theta_{\bx_{ij}}\nu) - \psi((1-\theta_{\bx_{ij}})\nu) ].
\end{aligned}
\end{equation*}
Thus the first term is 
{\small \begin{equation*}
\begin{aligned}
&\hspace{5mm}\frac{\partial^2 l_{ij}}{\partial\theta_{\bx_{ij}}^2}\cdot \left(\frac{\partial\theta_{\bx_{ij}}}{\partial\eta_{ij}} \right)^2\cdot \left(\frac{\partial\eta_{ij}}{\partial\bbet} \right)\left( \frac{\partial\eta_{ij}}{\partial\bbet}\right) ^\top\\
&=\nu^2[ \psi(\theta_{\bx_{ij}}\nu+t_{ij}) + \psi((1-\theta_{\bx_{ij}})\nu+y_{ij}-t_{ij} ) -\psi(\theta_{\bx_{ij}}\nu) - \psi((1-\theta_{\bx_{ij}})\nu) ]\theta^2_{\bx_{ij}}(1-\theta_{\bx_{ij}})^2\bx_{ij}\bx_{ij}^\top.
\end{aligned}
\end{equation*} }
The second term, which has expectation zero, is 
{\small \begin{equation*}
\begin{aligned}
&\hspace{5mm}\frac{\partial l_{ij}}{\partial\theta_{\bx_{ij}}}\cdot \frac{\partial^2\theta_{\bx_{ij}}}{\partial\eta_{ij}^2}\cdot \left(\frac{\partial\eta_{ij}}{\partial\bbet} \right)\left( \frac{\partial\eta_{ij}}{\partial\bbet}\right)^\top \\
&= \nu [ \phi(\theta_{\bx_{ij}}\nu+t_{ij}) - \phi((1-\theta_{\bx_{ij}})\nu+y_{ij}-t_{ij} ) -\phi(\theta_{\bx_{ij}}\nu) + \phi((1-\theta_{\bx_{ij}})\nu) ]  \theta_{\bx_{ij}} (1-\theta_{\bx_{ij}} )(1-2\theta_{\bx_{ij}})\bx_{ij}\bx_{ij}^\top.
\end{aligned}
\end{equation*} }
For each $i=1, 2$, $j=1,2,\ldots n_i$, let
\begin{equation*}
\begin{aligned}
a_{ij} &=  \phi(\theta_{\bx_{ij}}\nu+t_{ij}) - \phi((1-\theta_{\bx_{ij}})\nu+y_{ij}-t_{ij} ) -\phi(\theta_{\bx_{ij}}\nu) + \phi((1-\theta_{\bx_{ij}})\nu) \\
b_{ij} &=  \psi(\theta_{\bx_{ij}}\nu+t_{ij}) + \psi((1-\theta_{\bx_{ij}})\nu+y_{ij}-t_{ij} ) -\psi(\theta_{\bx_{ij}}\nu) - \psi((1-\theta_{\bx_{ij}})\nu).
\end{aligned}
\end{equation*}
Since the total log likelihood is $l=\sum_{i,j}l_{ij}$,
\begin{equation*}
\begin{aligned}
\frac{\partial l}{\partial\bbet} = \nu\sum\limits_{i,j} a_{ij} \theta_{\bx_{ij}} (1-\theta_{\bx_{ij}} )\bx_{ij} = \nu \bX^\top W_1\bz,
\end{aligned}
\end{equation*}
where the rows of $\bX$ are $\bx_{ij}^\top$, $W_1=\text{diag}(a_{ij})$ and $\bz=(\theta_{\bx_{11}} (1-\theta_{\bx_{11}} ),\ldots,\theta_{\bx_{1n_1}} (1-\theta_{\bx_{1n_1}} ),\ldots,\theta_{\bx_{2n_2}} (1-\theta_{\bx_{2n_2}} ))^\top$. The rows of $\bX, W_1$ and the elements of $\bz$ are ordered first by $j$ and then $i$.
\begin{equation*}
\begin{aligned}
\frac{\partial ^2l}{\partial\bbet\partial\bbet^\top} & = \nu^2\sum\limits_{i,j} b_{ij} \theta^2_{\bx_{ij}} (1-\theta_{\bx_{ij}} )^2 \bx_{ij}\bx_{ij}^\top + \nu\sum\limits_{i,j} a_{ij} \theta_{\bx_{ij}} (1-\theta_{\bx_{ij}} )(1-2\theta_{\bx_{ij}})\bx_{ij}\bx_{ij}^\top \\
& =-\nu\bX^\top W_2\bX,
\end{aligned}
\end{equation*}
where $W_2=-\text{diag}(\nu b_{ij}\theta^2_{\bx_{ij}} (1-\theta_{\bx_{ij}} )^2 + a_{ij} \theta_{\bx_{ij}} (1-\theta_{\bx_{ij}} )(1-2\theta_{\bx_{ij}})).$ The columns of $W_2$ is also ordered first by $j$ and then by $i$. 

When applying Laplace approximation to evaluate the marginal likelihood for a fixed $\nu$, 
$$
L_{\nu}=\int \exp\{l(\bbet) + \log \pi(\bbet) \}d\bbet,
$$
where $\pi$ is the prior on $\bbet$. For example, with $\pi(\bbet)$ is the independent normal $\text{N}(0,\sigma_k^2)$ on the $k$-th element of $\bbet$, let $h_{\nu}(\bbet)=l(\bbet) + \log \pi(\bbet)=l(\bbet) -\bbet^\top\Sigma^{-1}\bbet/2$, where $\Sigma=\mathrm{diag}(\sigma^2_1,\ldots, \sigma^2_d)$, we have
\begin{equation*}
\begin{aligned}
\frac{\partial h_{\nu}(\bbet)}{\partial\bbet} &=\frac{\partial l}{\partial\bbet} - \Sigma^{-1}\bbet \\
\frac{\partial^2 h_{\nu}(\bbet)}{\partial\bbet\partial\bbet^\top} &= \frac{\partial ^2l}{\partial\bbet\partial\bbet^\top} -\Sigma^{-1}.
\end{aligned}
\end{equation*}
Hence the Newton-Raphson step for solving the MLE of $\bbet$ given $\nu$ is given by
\begin{equation*}
\begin{aligned}
\hat\bbet^{(t+1)}=\hat\bbet^{(t)} + \left(\bX^\top W^{(t)}_2\bX+ \Sigma^{-1}/\nu\right)^{-1}\left( \bX^\top W_1^{(t)}\bz^{(t)}- \Sigma^{-1}\hat\bbet^{(t)}/\nu\right).
\end{aligned}
\end{equation*}

Under the alternative, suppose that $\pi(\bbet)=\pi(\bbet_1)\pi(\gamma)$, where $\bbet_1$ are the coefficients for the covariates and $\gamma$ for the group indicator. Instead of using independent normal prior on $\gamma$, the LIM $g$-prior \citep{li2015mixtures} could be adopted. Using the independent normal prior for $\bbet_1$, we have
\begin{equation*}
\begin{aligned}
h_{\nu}(\bbet)&=l(\bbet) + \log \pi(\bbet)\\
&=l(\bbet) -\bbet_1^\top\Sigma^{-1}\bbet_1/2 - g^{-1} \mathcal{J}_\nu(\hat\gamma)\gamma^2/2\\
&=l(\bbet) -\bbet_1^\top\Sigma^{-1}\bbet_1/2 - g^{-1}\nu (\bX^\top \hat W_2\bX)_2\gamma/2
\end{aligned}
\end{equation*}
where $(\bX^\top \hat W_2\bX)_2$ denote the block of the Hessian matrix corresponding to $\gamma$. Therefore,
\begin{equation*}
\begin{aligned}
\frac{\partial h_{\nu}(\bbet)}{\partial\bbet} &= \nu \bX^\top W_1\bz - \begin{pmatrix}\Sigma^{-1}\bbet_1  \\ g^{-1}\nu(\bX^\top\hat W_2\bX)_2\gamma \end{pmatrix}  \\
\frac{\partial^2 h_{\nu}(\bbet)}{\partial\bbet\partial\bbet^\top} &= -\nu\bX^\top W_2\bX-\begin{pmatrix}\Sigma^{-1} & \bzero \\ \bzero & g^{-1}\nu(\bX^\top \hat W_2\bX)_2 \end{pmatrix}.
\end{aligned}
\end{equation*}
The resulting NR update is
\begin{equation*}
\begin{aligned}
\hat\bbet^{(t+1)}=\hat\bbet^{(t)} + & \left(\bX^\top W^{(t)}_2\bX+ \begin{pmatrix} \Sigma^{-1}/\nu & \bzero \\ \bzero & g^{-1}(\bX^\top W_2^{(t)}\bX)_2 \end{pmatrix}\right)^{-1}\\
& \times \left( \bX^\top W_1^{(t)}\bz^{(t)}-\begin{pmatrix}\hat \Sigma^{-1}\bbet^{(t)}_1/\nu \\ g^{-1}(\bX^\top W^{(t)}_2\bX)_2\hat\gamma^{(t)} \end{pmatrix} \right).
\end{aligned}
\end{equation*}


\subsection{More on decision making}
\label{subsec:decmore}

We first consider the original hypothesis that there is no cross-group difference. Taking a decision theoretic perspective, let $d(\by)\in\{0,1\}$ be some decision rule, with $d(\by)=1$ corresponding to the rejection of the global null that there are no cross-group differences in the OTU composition. When the loss function is  
\begin{equation*}
\begin{aligned}
L(d(\by),c)=c\cdot\mathbbm{1}_{[H_0\text{ is true}]}d(\by) + (1-c)\cdot\mathbbm{1}_{[H_1\text{ is true}]}(1-d(\by)) \\
\end{aligned}
\end{equation*}
for some $0\leq c\leq 1$, one can show that the Bayes optimal decision rule is $d(\by)=\mathbbm{1}_{[\text{PJAP}>c]}.$ In particular, when $c=0.5$, this gives the optimal decision under the simple 0-1 loss. 

The decision on reporting the significant nodes is essentially a multiple testing problem. One way to address this problem is to use loss functions specified with the false positives and false negatives \citep{muller2006fdr}. For example, let $d_i(\by)\in \{0, 1\}$ be the decision rule on the $i$-th node; again, $d_i(\by)=1$ corresponds to the rejection of the node-specific null. Let FD and FN denote the number of false positives and false negatives. The posterior expectation of FD and FN are 
\begin{equation*}
\begin{aligned}
\overline{\text{FD}} &= \sum (1 - \text{PMAP}_i)\times d_i(\by),\\
\overline{\text{FN}} &= \sum \text{PMAP}_i \times (1-d_i(\by)).
\end{aligned}
\end{equation*}
It can be shown that under the loss $L(d(\by), t) = t\times  \text{FD} +  \text{FN}$, the Bayes optimal decision rule, which minimizes the posterior expected loss $\overline{L}(d(\by), t) = t\times \overline{\text{FD}} + \overline{\text{FN} }$ has the form $d_i(\by)=\mathbbm{1}_{[\text{PMAP}_i>c^\prime]}$ with the optimal threshold $c^\prime = t/(t+1)$, $t\geq 0$ \citep{muller2004optimal}. In our application, we use $c^\prime=0.5$ that corresponds to $t=1$ which is also recommended by \cite{barbieri2004optimal} from a Bayesian model choice perspective. {\color{black} Note that one can also consider loss functions that directly take into account the dependency among the hypotheses being tested. In our framework, such dependency is incorporated only through the probability model, not in the decision theoretic part.}


 
\subsection{Covariate selection}
\label{subsec:selection}

As we noted in Section~\ref{subsec:inference}, covariate selection is achievable in BGCR by putting a spike-and-slab prior on the regression coefficients \citep{george1997approaches}. For example, let $r_l \overset{\mathrm{ind}}{\sim} {\rm{Bernoulli}}(q_l)$ where $q_l\in (0, 1)$, $l=2,\ldots, p+1$. For $A\in\mI$, we can modify the prior on $\bbet(A)$ to be 
\begin{equation}
\begin{aligned}
\beta_l(A) \overset{\mathrm{ind}}{\sim} (1- r_l)\delta_0 + r_l {\rm{N}}(0, \sigma^2_l(A)), \quad l = 2, \ldots, p+1,
\end{aligned}
\label{eq:s1}
\end{equation}
where $\delta_0$ is a point mass at zero, $\sigma^2_l(A)$'s are chosen for ${\rm{N}}(0, \sigma^2_l(A))$ to cover all reasonable values of $\beta_l(A)$ while not supporting unreasonable values of $\beta_l(A)$. 

Let $\bm{r} = (r_2, \ldots, r_{p+1}) \in \{ 0, 1\}^p$. The independent Bernoulli priors on $r_l$ induce the following prior on $\bm{r}$
$$
\pi(\bm{r}) = \prod\limits^{p+1}_{l=2}q_l^{r_l}(1 - q_l)^{1 - r_l}.
$$
Conditioning on $\bm{r}$, the marginal likelihood of the data, $\phi_1(\Omega\mid \bm{r})$, is available as a byproduct of the BGCR inference algorithm (Section~\ref{subsec:inference}). When the number of covariates is not too large, this allows us to get the posterior of $\bm{r}$ by Bayes theorem:
\begin{equation*}
\begin{aligned}
\pi(\bm{r}\mid \bY) \propto \pi(\bm{r}) \phi_1(\Omega\mid \bm{r}).
\end{aligned}
\end{equation*}

We modify our simulation scenario $\RN{4}$ in Section~\ref{subsec:adj} to give a simple illustration of the covariate selection procedure. Consider the data simulated under the alternative, in which the counts of OTU `4481131' ($\omega_s$) are increased by $175\%$ in the second group. Instead of using ``gender'' as a confounder, we generate two covariates for each sample:
\begin{equation*}
\begin{aligned}
x_{ij2} \overset{\rm{iid}}{\sim} {\rm N}(0, 1),\quad x_{ij3} \overset{\rm{iid}}{\sim} {\rm N}(0, 1).
\end{aligned}
\end{equation*}
Suppose that the first covariate is relevant to the counts of a specific OTU while the second covariate has nothing to do with the OTU counts. Specifically, we increase the counts of OTU `4352657' ($\omega_c$) in the $j$-th sample in group $i$ by $(x_{ij2}\times 175\%)$ (when this value is less than $-1$, we set the count to zero). We note that due to the large variation in OTU counts, the signal injected on $\omega_c$ is quite weak.

Consider a specific round of simulation. We let $r_2 = r_3 = 0.5$, $\sigma^2_l(A)=10$ for $l=2, 3$ and fit BGCR with the prior in (\ref{eq:s1}). \ref{tab:prob} summaries the posterior probabilities of the four possible models. In comparison, each model has equal prior probabilities. Therefore, the important variable is correctly identified. 

\begin{table}[h]
\begin{center}
\begin{tabular}{c | c | c  | c |  c  }
\hline
Covariate in the model & None & 2 & 3 & 2 and 3 \\ \hline
Posterior probability  &  0.223 & 0.320 & 0.186  & 0.270 \\
\hline
\end{tabular}
\end{center}
\caption{Posterior probabilities of different models (no confounding).}
\label{tab:prob}
\end{table}

Although a covariate selection procedure can be incorporated in BGCR, one must proceed with caution since this can substantially affect or even invalidate the meaning of the testing result on the two-group difference. To see this intuitively, consider the following simplistic but representative scenario. Suppose there is a (close-to) perfect confounding covariate which explains virtually all the difference across the two groups. Once this covariate is included in the model then there is no remaining cross-group difference and the two-group comparison will not favor the alternative. However, including the covariate into the model may not improve the fit to the observed data in any substantive manner as its effect is largely overlapping with that of the intercept (i.e., the group label). Consequently, statistical model selection strategies, both Bayesian or frequentist, would very likely to exclude this covariate from the model. This would lead to a significant testing result on the two-group differences. As a simple illustration, in the previous example, suppose instead we have
$$
x_{1j2} \overset{\rm{iid}}{\sim} {\rm N}(0, 1),\quad x_{2j2} \overset{\rm{iid}}{\sim} {\rm N}(2, 1).
$$
In this case, the first covariate is a confounding variable. \ref{tab:prob2} summaries the posterior probabilities of the four possible models. Due to the strong confounding effect, the first covariate is excluded from the model, which would lead to false positives in the testing scenario.

\begin{table}[h]
\begin{center}
\begin{tabular}{c | c | c  | c |  c  }
\hline
Covariate in the model & None & 2 & 3 & 2 and 3 \\ \hline
Posterior probability  &  0.599 & 0.001 & 0.400  & $\approx 0$ \\
\hline
\end{tabular}
\end{center}
\caption{Posterior probabilities of different models (with confounding).}
\label{tab:prob2}
\end{table}



 \clearpage

\subsection{Additional figures }
 \vfill
  
\begin{figure}[!h]
\begin{center}
\includegraphics[width = 16.5cm]{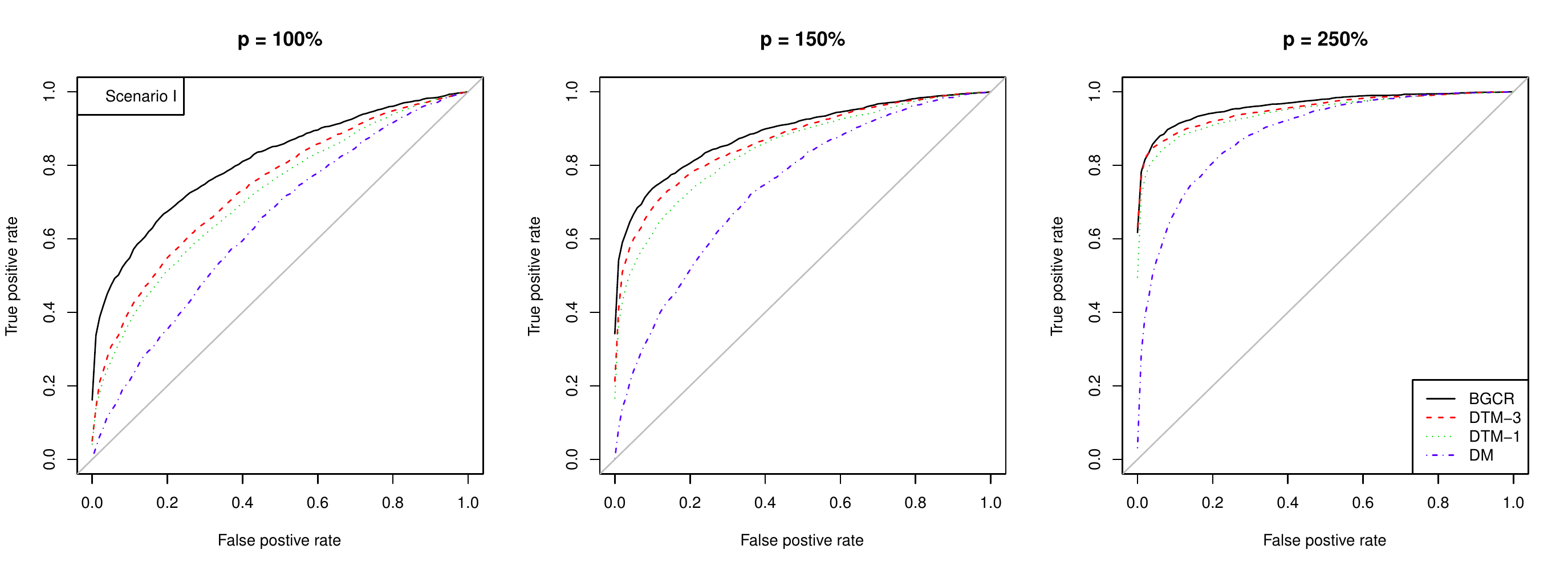}\\
\includegraphics[width = 16.5cm]{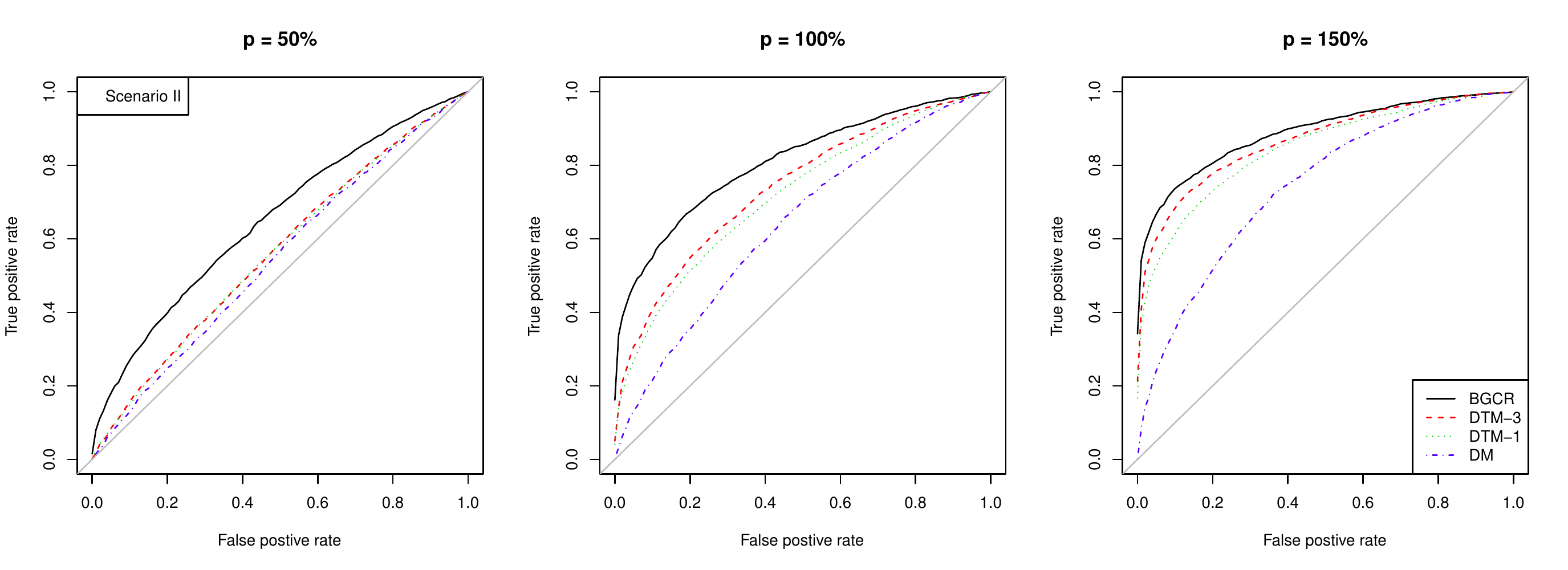}\\
\caption{ROC curves for Scenario $\RN{1}$ and $\RN{2}$ with $K = 50$. The columns are indicated by the percent of count increased in the second group ($p$). }
\label{fig:roc_2}
\end{center}
\end{figure}
\vfill

\begin{figure}[!h]
\begin{center}
\includegraphics[width = 16.5cm]{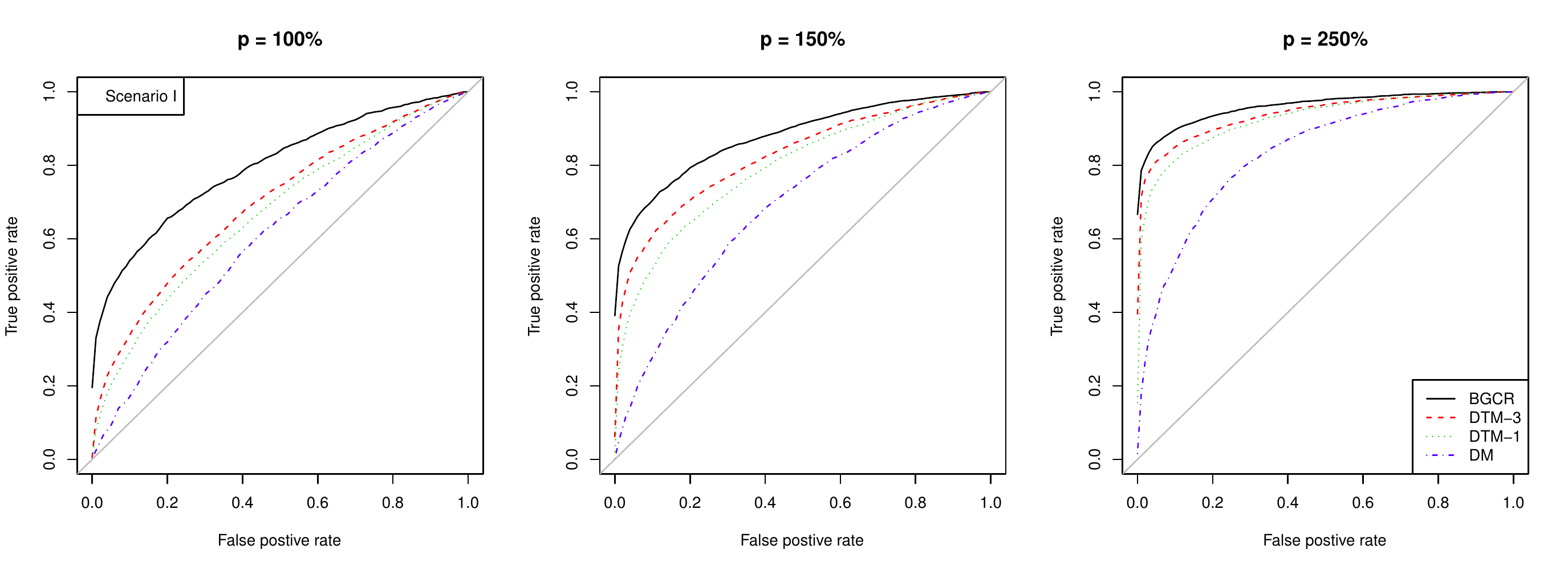}\\
\includegraphics[width = 16.5cm]{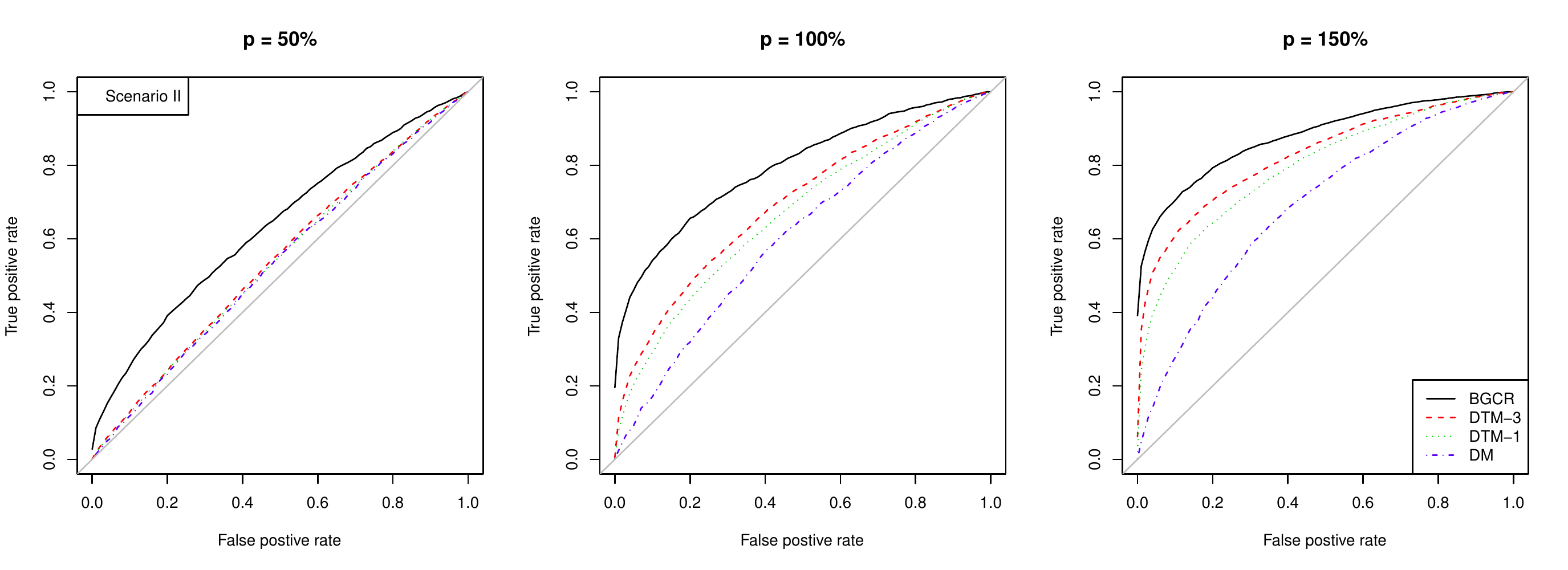}\\
\caption{ROC curves for Scenario $\RN{1}$ and $\RN{2}$ with $K = 75$. The columns are indicated by the percent of count increased in the second group ($p$). }
\label{fig:roc_3}
\end{center}
\end{figure}

\begin{figure}[ht]
\begin{center}
\includegraphics[width = 16.5cm]{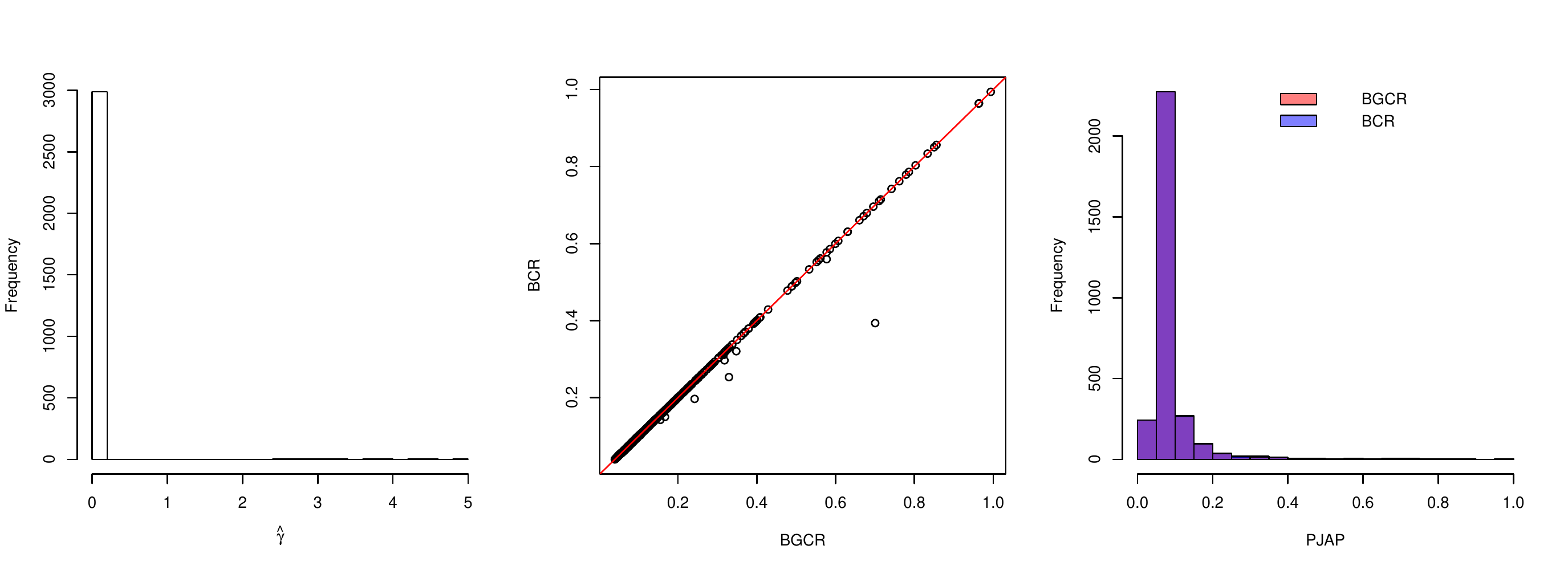}\\
\caption{BGCR vs BCR under the null in scenario 3. Left: Histogram of the estimated $\gamma$ in BGCR; Middle: PJAPs of BGCR vs BCR; Right: Histograms of the PJAPs of BGCR and BCR.}
\label{fig:sim3_null}
\end{center}
\end{figure}

\begin{figure}[!h]
\begin{center}
\includegraphics[width = 16.5cm]{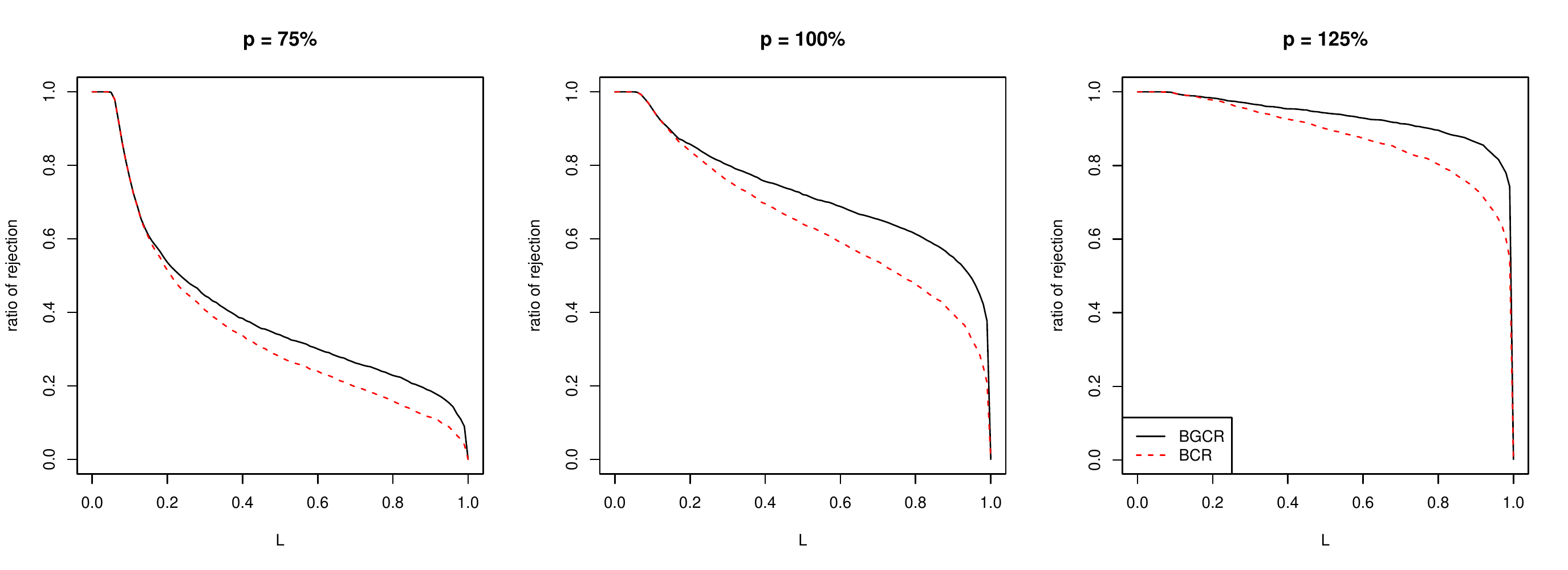}\\
\caption{Ratio of rejection under the alternatives in Scenario $\RN{3}$. {\color{black} The columns are indicated by the percent of count increased in the second group ($p$).}}
\label{fig:sim3_power}
\end{center}
\end{figure}

\begin{figure}[!ht]
\begin{center}
\includegraphics[width = 16.5cm]{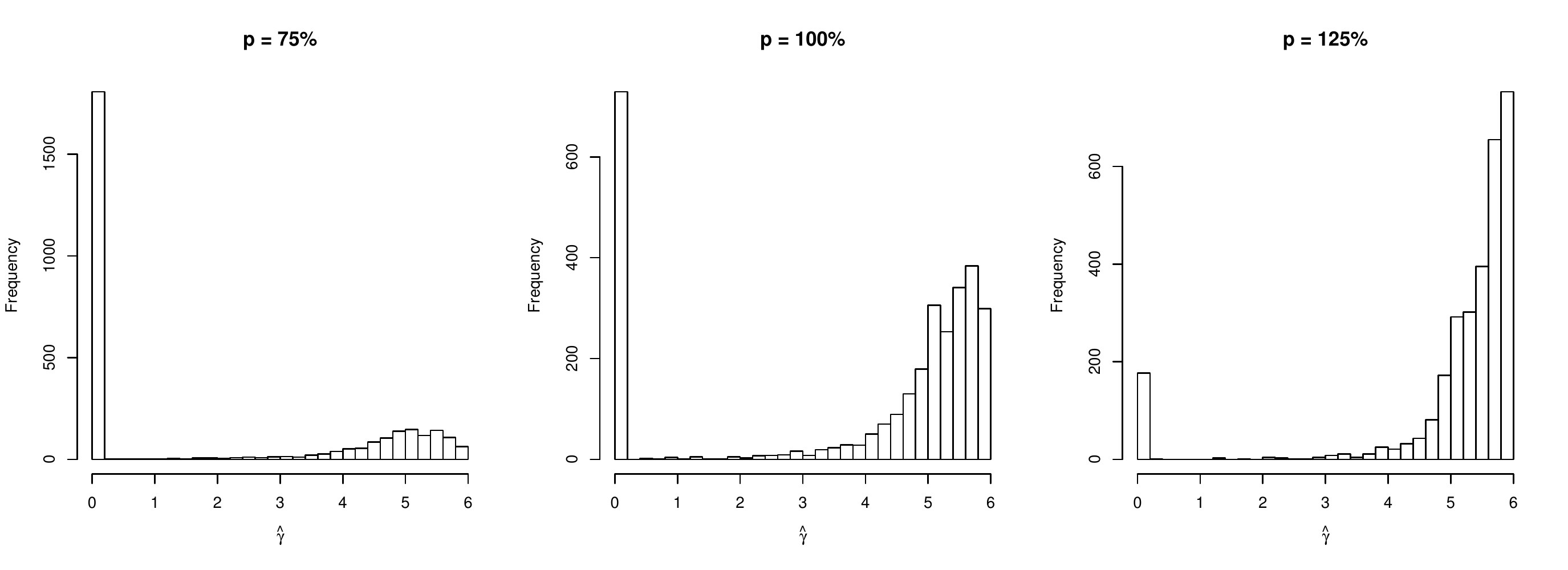}\\
\caption{Estimated $\gamma$ under the alternatives in scenario 3.}
\label{fig:sim3_gamma}
\end{center}
\end{figure}

\begin{figure}[!ht]
\begin{center}
\includegraphics[width = 17cm]{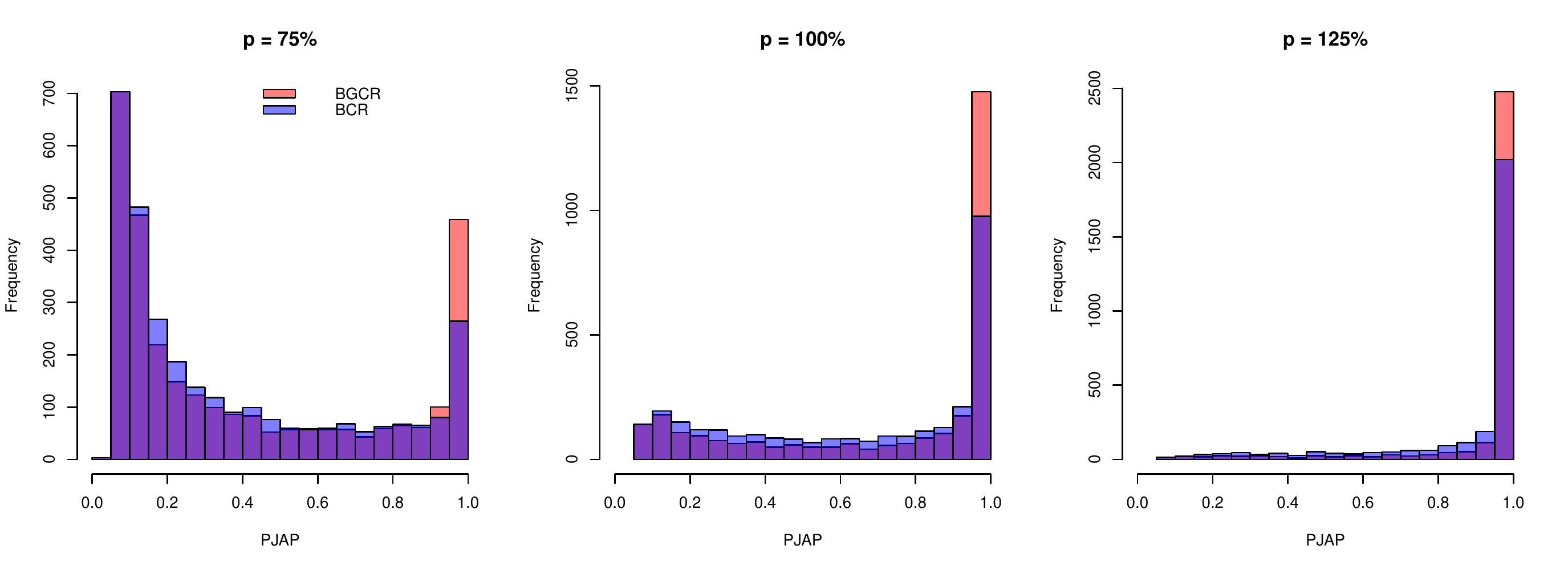}\\
\caption{Histograms of the PJAPs under the alternatives in scenario 3. }
\label{fig:sim3_alt}
\end{center}
\end{figure}

\begin{figure}[!h]
\begin{center}
\includegraphics[width = 14cm]{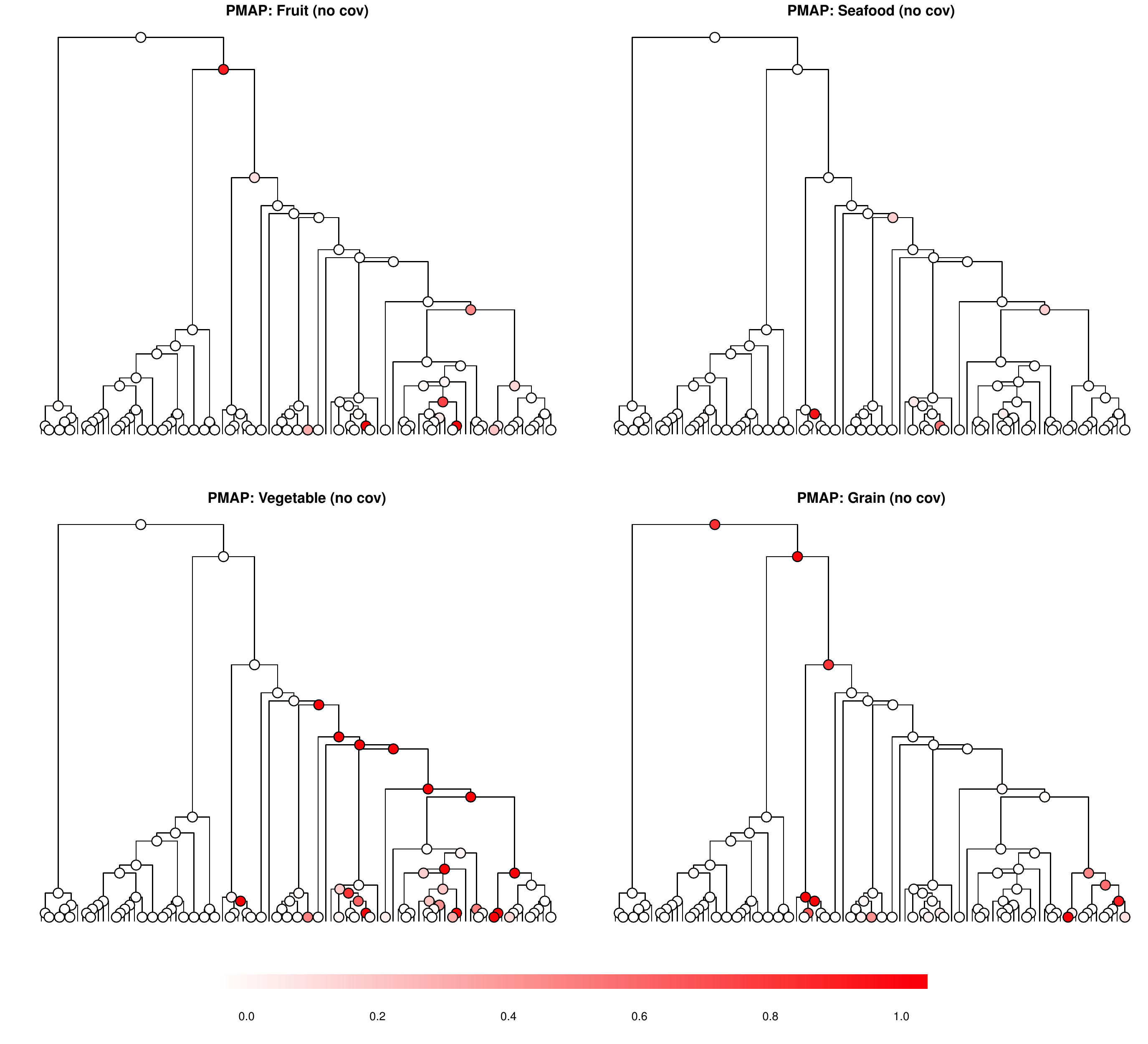}
\caption{PMAPs for the four comparisons that reject the global null. The nodes are colored by PMAPs reported by BGCR with no covariate adjusted.}
\label{fig:app_pmap2}
\end{center}
\end{figure}
 
 \begin{figure}[!h]
\begin{center}
\includegraphics[width = 14cm]{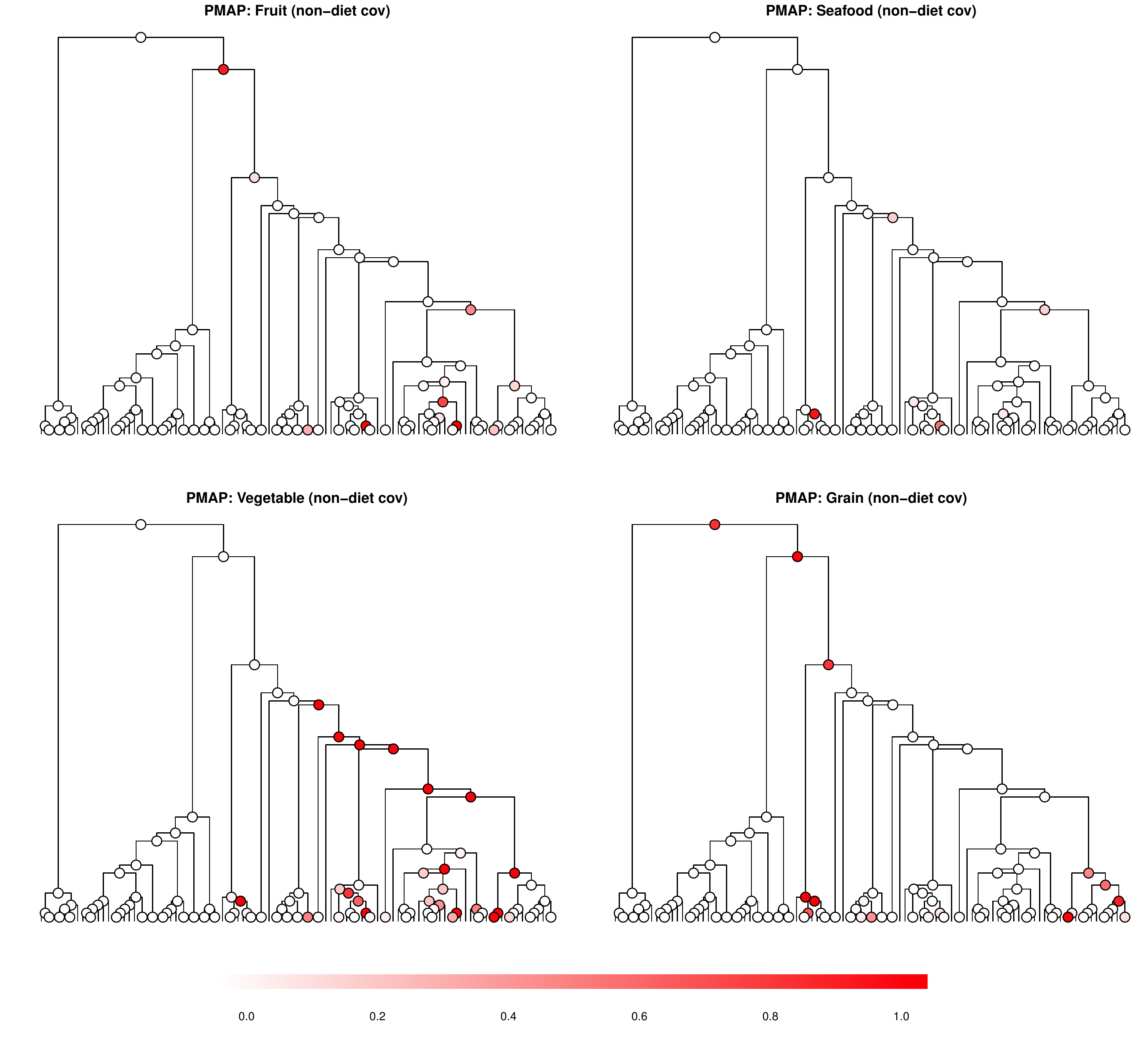}
\caption{PMAPs for the four comparisons that reject the global null. The nodes are colored by PMAPs reported by BGCR with only non-dietary covariates adjusted.}
\label{fig:app_pmap3}
\end{center}
\end{figure}
 
  \begin{figure}[!h]
\begin{center}
\includegraphics[width = 14cm]{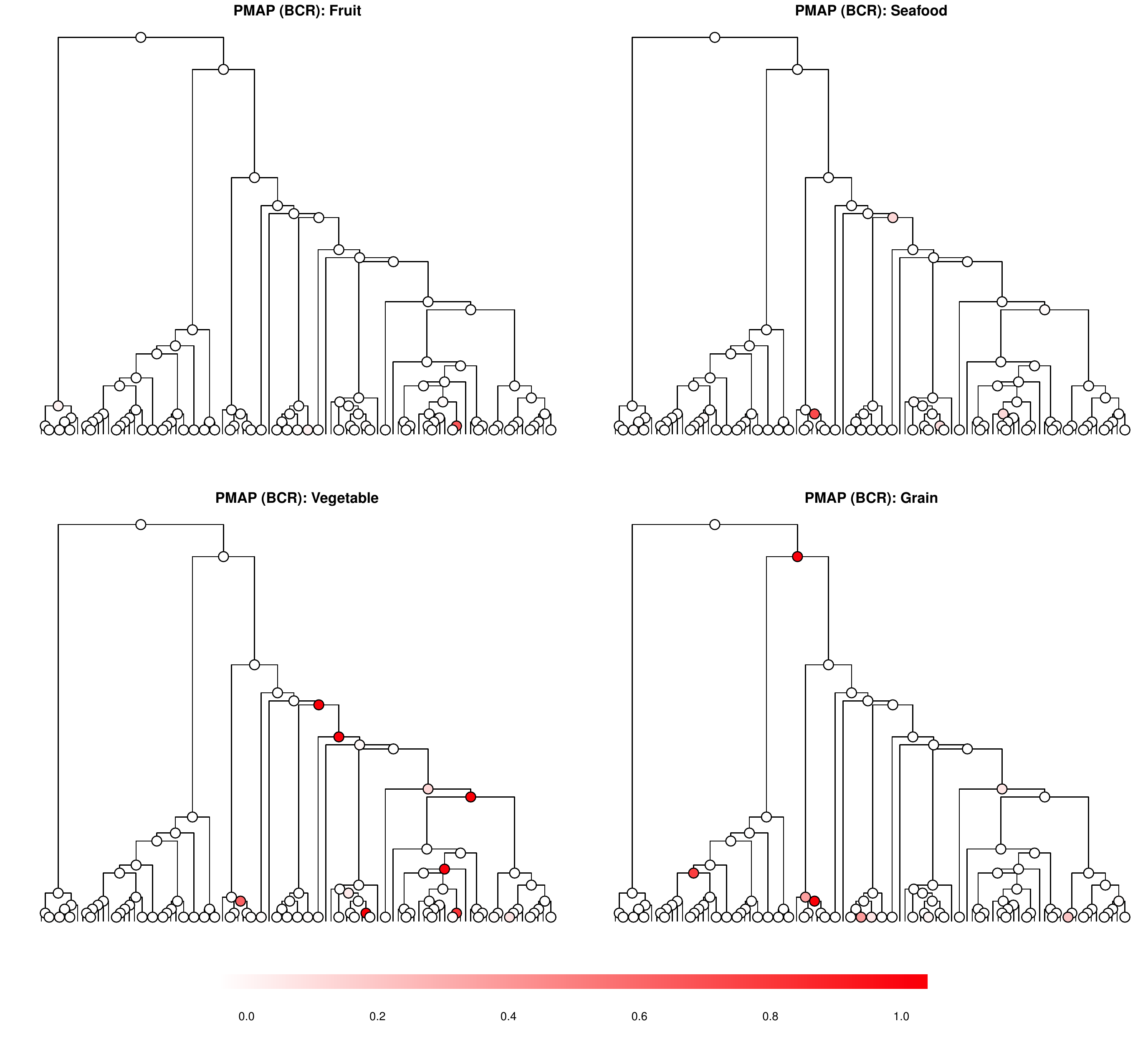}
\caption{PMAPs for the four comparisons that reject the global null. The nodes are colored by PMAPs reported by {\color{black}BCR} with both non-dietary covariates and dietary covariates adjusted.}
\label{fig:app_pmap_ind}
\end{center}
\end{figure}

\end{document}